\documentclass[traditabstract]{aa}  
\usepackage{graphicx}

\usepackage{txfonts}

\usepackage{xcolor}
\usepackage{multirow}
\usepackage{siunitx}
\usepackage{verbatim}
\usepackage{bm}
\usepackage{soul}

\newcommand{\rey}{\mathcal{R}}
\def\mstar  {$M_{\star}$}

\newcommand{\degree}{\ensuremath{^\circ}}

\def\arcsec{\hbox{$^{\prime\prime}$}}
\newcommand{\rdust}{{$R_{\rm disc,dust}$}}

\newcommand{\rgas}{{$R_{\rm disc,gas}$}}

\begin{document}

   \title{Observational constraints on disc sizes \\ in protoplanetary discs in multiple systems in the Taurus region}

   \subtitle{II. Gas disc sizes}

   \author{A.A. Rota\inst{\ref{instMI},\ref{instESO}}
          \and
          C.F. Manara \inst{\ref{instESO}}
	\and A. Miotello\inst{\ref{instESO}} \and G. Lodato\inst{\ref{instMI}} \and S. Facchini\inst{\ref{instESO}} \and M. Koutoulaki\inst{\ref{instESO}} \and  G. Herczeg\inst{\ref{instKIAA}} \and F. Long\inst{\ref{instCfA}} \and M. Tazzari\inst{\ref{instIoA}} \and S. Cabrit\inst{\ref{instLERMA}} \and D. Harsono\inst{\ref{instASIAA},\ref{instNTHU}} \and F. M\'enard\inst{\ref{instGRE}} \and P. Pinilla\inst{\ref{instHD}, \ref{instUK}} \and G. van der Plas\inst{\ref{instGRE}} \and E. Ragusa\inst{\ref{instLei},\ref{instENS}}  \and Hsi-Wei Yen\inst{\ref{instASIAA}}
          }

   \institute{Universit\`a degli Studi di Milano, Via Celoria 16, 20133 Milano, Italy\label{instMI}\\
              \email{alessiaannie.rota@studenti.unimi.it}
         \and
            European Southern Observatory, Karl-Schwarzschild-Strasse 2, 85748 Garching bei M\"unchen, Germany\label{instESO}\\
              \email{cmanara@eso.org}
\and  Kavli Institute for Astronomy and Astrophysics, Peking University, Yiheyuan 5, Haidian Qu, 100871 Beijing, China
\label{instKIAA}
\and Center for Astrophysics \textbar\, Harvard \& Smithsonian, 60 Garden St., Cambridge, MA 02138, USA \label{instCfA}
\and
            Institute of Astronomy, University of Cambridge, Madingley Road, CB3 0HA Cambridge, UK\label{instIoA}
\and
Observatoire de Paris, PSL University, Sorbonne University, CNRS, LERMA, 61 Av. de l'Observatoire, 75014 Paris, France\label{instLERMA}
        \and
            Academia Sinica Institute of Astronomy and Astrophysics, 11F of Astro-Math Bldg, 1, Sec. 4, Roosevelt Rd, Taipei 10617, Taiwan\label{instASIAA}
\and
Institute of Astronomy, Department of Physics, National Tsing Hua
University, Hsinchu, Taiwan\label{instNTHU}
\and
            Univ. Grenoble Alpes, CNRS, IPAG, F-38000 Grenoble, France\label{instGRE}
\and        Max-Planck-Institut f\"{u}r Astronomie, K\"{o}nigstuhl 17, 69117, Heidelberg, Germany. \label{instHD}
\and        Mullard Space Science Laboratory, University College London, Holmbury St Mary, Dorking, Surrey RH5 6NT, UK. \label{instUK}
\and
            School of Physics and Astronomy, University of Leicester, Leicester LE1 7RH, UK\label{instLei}
\and
            Univ Lyon, Univ Lyon1, Ens de Lyon, CNRS, Centre de Recherche Astrophysique de Lyon UMR5574, F-69230, Saint-Genis-Laval, France\label{instENS}
}

   \date{Received April 9, 2021; accepted -}

  \abstract{The formation of multiple stellar systems is a natural by-product of the star-formation process, and its impact on the properties of protoplanetary discs and on the formation of planets is still to be fully understood. 
To date, no detailed uniform study of the gas emission from a sample of protoplanetary discs around multiple stellar systems has been performed. Here we analyse new ALMA observations at a $\sim$21 au resolution of the molecular CO gas emission targeting discs in eight multiple stellar systems in the Taurus star-forming regions. $^{12}$CO gas emission is detected around all primaries and in seven companions. With these data, we estimate the inclination and the position angle for all primary discs and for five secondary or tertiary discs, and measure the gas disc radii of these objects with a cumulative flux technique on the spatially resolved zeroth moment images.  
When considering the radius including 95\% of the flux as a metric, the estimated gas disc size in multiple stellar systems is found to be on average $\sim 4.2$ times larger than the dust disc size. This ratio is higher than what was recently found in a population of more isolated and single systems. On the contrary, when considering the radius including 68\% of the flux, no difference between multiple and single discs is found in the distribution of ratios. This discrepancy is due to the sharp truncation of the outer dusty disc observed in multiple stellar systems.
The measured gas disc sizes are consistent with tidal truncation models in multiple stellar systems assuming eccentricities of $\sim0.15$-$0.5$, as expected in typical binary systems. 
  }

   \maketitle

\section{Introduction}

The collapse and fragmentation of molecular cloud cores frequently leads to the formation of binary stars or higher order multiple systems \citep[e.g.,][]{2018Bate,2014Reipurth+}. In a tight multiple system, the outer parts of the individual discs in the system are affected by tidal forces from the companions, usually resulting in disc truncation. 
 The effect of tidal disruption of discs on the ability of protoplanetary discs in multiple systems to form planets is still a matter of investigation.

From an observational point of view, exoplanet detection surveys were initially strongly biased towards binary systems with separation  $\ge 200$ au \citep{2010Eggenberger_Udry}, because these searches focused on stellar environments as similar as possible to the solar system. However, in 2003 the first exoplanet in a close binary was detected in the $\gamma$ Cephei system \citep{2003Hatzes+} and exoplanets have been detected in many multiple systems \citep{2016Hatzes}, also in close-binaries with separation $\sim 20 $ AU. 
In particular, 217 exoplanets are known in binary-star systems and 51  in higher multiplicity systems (updated to Sep 2021)\footnote{The data of all detected planets are collected in the Exoplanet-catalogue maintained by J. Schneider (http://exoplanet.eu); whereas the binary and higher multiplicity systems can be found separately in the catalogue of exoplanets in binary star systems maintained by R. Schwarz (\citealp{2016Schwarz+}; http://www.univie.ac.at/adg/schwarz/multiple.html).}.

In multiple systems, dynamical interactions experienced by circumstellar discs affect their evolution \citep{1977Papaloizou_Pringle,1994Artymowicz_Lubow,2018Rosotti_Clarke,zagaria21a} and may have a negative impact on the formation and evolution of planetary systems. Tidal effects due to interactions may lead the outer parts of the discs to become strongly misaligned or warped \citep{2013Lodato_Facchini}, impacting planet migration and orbit stability and evolution (\citealp{1999Holman_Wiegert}; \citealp{2014Jensen_Akeson}). 
Moreover, disc truncation may affect the availability of the building blocks of planets \citep{2011Zsom_Dullemond, 2016Kraus+} and may reduce disc lifetimes of either or both the disc around the primary or secondary component \citep{2007Monin+}. These considerations could suggest that multiplicity reduces the probability of planet formation, since there is insufficient mass to form planets or insufficient time for planetesimal formation to operate before the disc is dispersed.

The detection of planets in multiple stellar systems and in close-binaries has triggered studies investigating how such planets could form and, more generally, about how planet-formation is affected by binarity \citep{2015Thebault_Haghipour}. In particular, to investigate these two aspects, the observation of circumstellar discs in multiple systems is crucial to study how discs evolve and how interactions between discs in multiple systems affect their evolution.

Surveys of multiple stellar systems conducted with the Submillimeter Array (SMA) and the Atacama Large
Millimeter/submillimeter Array (ALMA) are providing
constraints on the theory of disc evolution in multiple systems. In particular, observations of discs around multiple stars in the Taurus and $\rho$-Ophiucus regions (\citealp{2012Harris+, 2014Akeson_Jensen, 2017Cox+, 2019Akeson+, 2019Manara+,2019Long+, 2021Zurlo+}) show that discs in multiple systems are on average fainter in the (sub)millimeter at any given stellar mass than those in single systems. The dusty disc around the primary component, the more massive star, is usually brighter and more massive than the disc around the secondary component. 

\cite{2019Manara+} discussed a sample of ten multiple stellar systems in the Taurus star-forming region, composed by eight binaries and two higher-order multiple systems $-$ UZ Tau (e.g., \citealp{2001White_Ghez}) and T Tau (e.g., \citealp{2016Kohler+}).
The sample was part of a high angular resolution ($\sim 0\farcs12$) and high sensitivity ($\sim 50 ~\si{\micro}$Jy beam$^{-1}$ at 225 GHz - 1.33 mm) survey of 32 protoplanetary discs around stars with spectral type earlier than M3 \citep{2018Long+,2019Long+} and covers a wide range of system parameters.
\cite{2019Manara+} showed that, in general,
the dust radii of the discs around secondary or tertiary components of multiple systems are statistically smaller than the sizes of the discs around primary stars in multiple systems, which are in turn smaller than the discs around single stars and whose brightness profiles present a steeper outer edge than singles. 
The typical measured ratio between $R_\mathrm{disc,dust}$ and the projected separation ($a_\mathrm{p}$) was $\lesssim 0.1$. 
This result can be reconciled with tidal truncation models \citep[e.g.,][]{1977Papaloizou_Pringle,1994Artymowicz_Lubow} only in the highly improbable case that the observed binary systems were all on highly eccentric ($e>$ 0.7) orbits. Assuming that dust radii are smaller than gas radii by factors $\gtrsim 2-3$ \citep[e.g.,][]{2018Ansdell+}, possibly due to a more effective drift of the dust probed by 1.3 mm observations, would result in values of eccentricities in agreement with the expected ones, $0\lesssim e \lesssim 0.5$ \citep[e.g.,][]{2013Duchene_Kraus}. 
However, the analysis above was performed using the dust disc radii estimates, since the data available did not cover the gas emission in these discs, which is however the component that we need to compare with current models of tidal truncation as this is the one that better responds to the disc dynamics.
By considering the effect of dust growth and drift in truncated systems, \citet{zagaria21b} recently showed that the results by \citet{2019Manara+} can be reconciled with tidal truncation models without the need to invoke extremely high eccentric orbits. This result must however be confirmed also observationally by measuring the gas size of discs around multiple systems.

In this work, we analyse new high-resolution ALMA observations of line and continuum emission in the same sample of multiple stellar systems in the Taurus star-forming region as of \citet{2019Manara+}. Our aim is to test tidal truncation models by measuring for the first time the gas size of discs in multiple stellar systems in a uniform way, with the same instrument and a similar noise level at high spatial resolution.

The paper is organised as follows. In Sect.~\ref{sect:sample} we present the sample analysed in our work. In Sect.~\ref{sect:alma_reduction} we discuss the observations, and how the data were calibrated. 
In Sect.~\ref{sect:analysis} we analyse the new ALMA observations, deriving the geometrical properties of the discs and the disc radii. We then discuss the results in Sect.~\ref{sect:discussion}, comparing the measured radii to theoretical models to constrain the eccentricities of the orbits expected with our data. Finally, we draw our conclusions in Sect.~\ref{sect:conclusions}.

\begin{figure*}[]
    	\centering
	    \includegraphics[width=\textwidth,keepaspectratio]{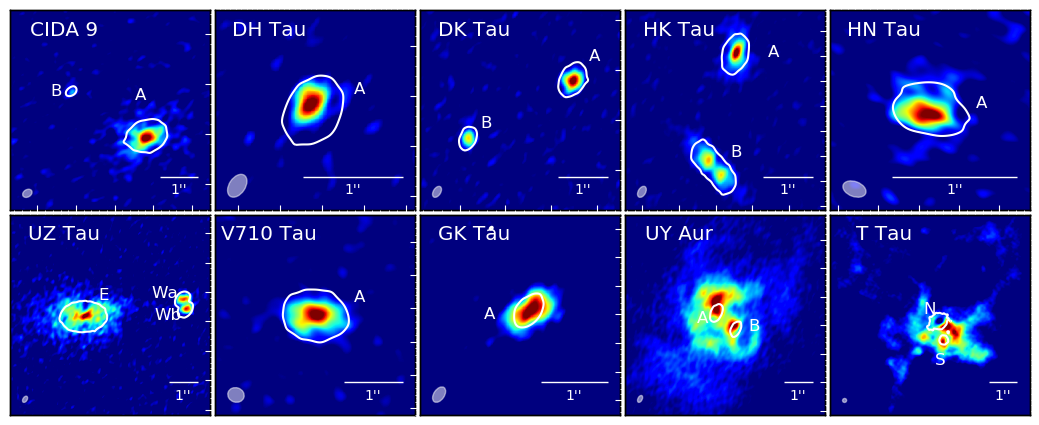}
	\caption{$^{12}$CO zeroth moment images of the discs around multiple stars in the Taurus star-forming region observed in the new ALMA observations. All bars in the bottom right of each panel are 1\arcsec long, which is $\sim 140$ au at the typical distance of the Taurus region. The FWHM beam size is shown in the bottom left of each panel.  In each image, the color scale has the peak flux as the maximum, and two times the image rms as minimum. White contours show 5 times the rms of the continuum emission. The components of the systems are labelled. The label for any undetected secondary components in the continuum emission is not shown.  }\label{fig:mysampleMOM0}
\end{figure*}

\section{Sample properties}\label{sect:sample}

The sample of targets analysed in this work comprises eight targets. Seven binary systems -- CIDA 9, DH Tau, DK Tau, HK Tau, HN Tau, RW Aur, V710~Tau, and one quadruple system -- UZ~Tau. The primary components of the CIDA 9 and UZ Tau systems are very bright, and their circumprimary discs show ring-like structures, while the other systems have smooth discs around the primary stars  \citep{2018Long+}.

Eleven systems were originally selected for the apparent multiplicity of their stars (10 systems from \citealt{2019Manara+} and GK Tau from \citealt{2019Long+}). New ALMA observations were requested to detect CO lines in nine targets out of ten of \cite{2019Manara+} (i.e. CIDA 9, DH Tau, DK Tau, HK Tau, HN Tau, T Tau, UY Aur, UZ Tau, V710 Tau) and in GK Tau. For the last target of \citet{2019Manara+}, RW Aur, CO observations with similar resolution and sensitivity were already available in the ALMA archive \citep{2018Rodriguez+}. 
Three sources -- GK Tau, UY Aur, and T Tau -- were then excluded from the analysis performed in this work.
GK Tau and GI Tau are possibly part of a bound system. However, considering the high separation of the targets ($\sim 1700$ au; \citealp{1994Hartigan+}, \citealp{HH14})) the sizes of both GI Tau and GK Tau are not affected by tidal truncation \citep{2020Pearce+}, and are thus excluded from the analysis in this work.
Finally, the observation of UY Aur and T Tau do not show a velocity pattern compatible with Keplerian rotation (see Section \ref{sect:alma_reduction}), making it difficult to determine if the gas emission is reflecting the disc dynamics, and therefore to estimate the radius of individual discs. The data for these two objects will be analysed in future work.

Table \ref{tab:ch4paramDisc} reports information on the eight multiple stellar systems analysed in this work. 
We assumed the stellar masses inferred by \cite{2019Manara+}. In that work, the effective temperature was used with the stellar luminosity corrected assuming the Gaia DR2 distances of the targets \citep{2018GaiaColl}, and masses were obtained assuming the evolutionary models by \cite{2015Baraffe+} and the nonmagnetic models by \cite{2016Feiden}, as already done in \cite{2016Pascucci+} and \citet{2019Long+}. Only in two cases the stellar masses are derived from orbital dynamics and not as just described: UZ Tau E \citep{2000Simon+} and HN Tau A \citep{2017Simon+}.

The projected separations ($a_\mathrm{p}$, assumed from \citealp[][and references therein]{2019Manara+}) of the individual components in the systems range from $\sim 49$ au to 461 au, the mass ratios ($q=M_2/M_1$) from $\sim 0.9$ to $\sim 0.1$, and the mass parameters ($\mu=M_1M_2/(M_1+M_2)$) from $\sim 0.1$ to $\sim 0.5$.

\begin{table*} 
\centering 
\caption{Information on the stellar and multiplicity properties of the targets analysed in this work}
\begin{tabular}{l*{9}{c}}

 &   $a_\mathrm{p}$ [\arcsec] &     $d$ [pc] & SpT1 &   SpT2 & $M_1$ [$M_\odot$] &    $M_2$ [$M_\odot$] &       $q$ & $\mu$  \\
 
\hline\hline

CIDA 9 & 2.35 & 171 &  \tiny{M2} & \tiny{M4.5} & 0.43$^{+0.15}_{-0.10}$ & 0.19$\pm$0.1 & 0.44 & 0.31  \\
DH Tau & 2.34 & 135 & \tiny{M2.5} &  \tiny{M7.5} & 0.37$^{+0.13}_{-0.10}$ &  0.04$\pm$0.2 &  0.11 &  0.10   \\
DK Tau & 2.38 & 128 & \tiny{K8.5} & \tiny{M1.5} & 0.60$^{+0.16}_{-0.13}$ & 0.44$\pm$0.2 & 0.73 & 0.42  \\
HK Tau  & 2.32 & 133 & \tiny{M1.5} & \tiny{M2} & 0.44$^{+0.14}_{-0.11}$ & 0.37$\pm$0.2 & 0.84 & 0.46  \\
HN Tau & 3.16 & 136 & \tiny{K3} & \tiny{M5} & 1.53$\pm$0.15 & 0.16$\pm$0.1 & 0.10 & 0.09  \\
RW Aur & 1.49 & 163 & \tiny{K0} & \tiny{K6.5} & 1.20$^{+0.18}_{-0.13}$ & $0.81\pm0.2$ & 0.67 & 0.41  \\
UZ Tau$^{*}$& 3.52 & 131 & \tiny{M2} & \tiny{M3} & 1.23$\pm$0.07
& 0.58$\pm$0.2 & 0.47 & 0.32  \\
UZ Tau W & 0.375 & 131 & \tiny{M3} & \tiny{M3} & 0.30$\pm$0.04 & 0.28$\pm$0.2 & 0.93 & 0.48  \\
V710 Tau & 3.22 & 142 & \tiny{M2} & \tiny{M3.5} & 0.42$^{+0.13}_{-0.11}$ & 0.25$\pm$0.1 & 0.60 & 0.37  \\

\hline
\end{tabular}
\tablefoot{ We report the projected separations $a_\mathrm{p}$, the distance $d$ from the observer, the spectral type of the primary (SpT1) and of the secondary (SpT2) stars, the masses of the primary and the secondary stars ($M_1$ and $M_2$ respectively) the mass ratios $q$, and the mass parameters $\mu$. $^{*}$ For UZ Tau, $M_{1}$ refers to the total mass of the UZ Tau E spectroscopic binary, while $M_{2}$ is the sum of the masses of UZ Tau Wa and UZ Tau Wb.}
\label{tab:ch4paramDisc}
\end{table*}

\section{Observations and data reduction}\label{sect:alma_reduction}
The ALMA observations presented here were obtained in program 2018.1.00771.S, (PI: Manara).
The observations were conducted in ALMA Band 6 with an angular resolution of $0\farcs15$ $- \sim 21$ au at the distance of Taurus, and with an integration time of $\sim 40$ min per source. These observations present an angular resolution slightly lower and a longer integration time than the previous Band 6 observations presented by \citet{2018Long+,2019Long+} and \citet{2019Manara+}, i.e. $0\farcs12$ and $\sim 4-9$ min/source, respectively.
In fact, the aim of the new observations was to detect different molecular emission lines with a higher signal-to-noise (SNR). The spectral set-up was chosen to include one continuum band (centered at 233 GHz), and the following molecular emission lines: $^{12}$CO $(J=2-1)$, $^{13}$CO $(J=2-1)$, C$^{18}$O $(J=2-1)$, N$_2$D $(J=3-2)$, DCN $(v =0, J=3-2)$, and H$_2$CO  $3(0,3)-2(0,2)$. Two configurations were requested, C43-6 and C43-3, but only observations in the former configuration were executed. Therefore, the maximum recoverable scale of the observations was only $\sim$1.8\arcsec. 

The targets were observed in two Scheduling Blocks (SBs), one including HN~Tau, T~Tau, and V710~Tau, and the other one including the remaining seven targets of the program. Each science target was observed multiple times in different Execution Blocks (EBs), three and six for each SB respectively.
The observations of the first SB, including three science targets, were executed between August 17, 2019 and September 19, 2019, while those of the second SB, including seven targets, were carried out between September 21, 2019 and September 27, 2019. J0510+1800 was used as flux, bandpass, pointing, and atmosphere calibrator, while J0440+1437 and J0438+3004 were used as phase calibrators. 

The calibrated data were retrieved using the script for PI provided by the observatory. In order to improve the SNR, the data were self-calibrated, in each EB separately. This step was performed using the Common Astronomy Software Applications package \citep[CASA, version 5.6.1,][]{2007McMullin+}.
To speed up the calibration, we applied a channel averaging of 125MHz in each spectral window (spw): this led to 8 channels in the continuum spw, centered on 233.0 GHz with a bandwidth of 1875.00 MHz, and one channel in the other five spws, used for the line observations. The channel averaging was done 
after flagging any channel in which emission lines were found.
In each EB we applied from one to four rounds of phase-only self-calibration, stopping the calibration when no significant improvement was measured in the SNR from one step to the next one. We then concatenated all EBs and applied one round of amplitude self-calibration. The final improvement in SNR has always been larger than $100\%$ for all targets, thanks to both a drop in the noise and an improvement in the observed flux.
As final step, we have applied the phase and amplitude calibration to the spectral windows containing emission lines.

Continuum and line images were created with the CASA task \textit{tclean}. Continuum images were created using both the continuum spectral window and the line-free channels of the three spws in which CO isotopologues emission were detected.
The cleaning was performed with Briggs weighting and a robust parameter of $+0.5$ using an interactive cleaning, except for the line images of HN Tau, GK Tau, UY Aur and T Tau that were created through the CASA \textit{auto-multithresh} algorithm with Briggs weighting and a robust parameter of $+0.5$. The cleaning was performed with a channel width of $0.1$ km/s for $^{12}$CO emission, except in the case of HN Tau where a $0.2$ km/s channel width was used. The $^{13}$CO and C$^{18}$O lines were cleaned with channel width of $0.2-0.5$ km/s.

\begin{table}[h]\centering\small
\caption{Information on the cleaned images}
	\begin{tabular}{cccc}
		Target   & & rms & Beam size\\
		   & & [mJy/beam per channel] & [$\arcsec \times \arcsec$] \\
		\hline\hline
		CIDA 9 & \multirow{3}{*}{}continuum  & 3.0e-02             &    $0.25 \times 0.19$   \\
	                                        &$^{12}$CO  & 	4.9                 &    $0.27 \times 0.20$    \\
	                                        &$^{13}$CO  & 	5.0                 &    $0.28 \times 0.21$    \\
	                                        &C$^{18}$O  & 	2.3                 &    $0.28 \times 0.21$   \\
	    \hline
	    DH Tau   & \multirow{3}{*}{}continuum  & 	2.7e-02                &    $0.23 \times 0.15$    \\
	                                        &$^{12}$CO  & 	 4.6                &    $0.25 \times 0.16$     \\
	                                        &$^{13}$CO  & 	3.6                 &    $0.27 \times 0.17$   \\
	                                        &C$^{18}$O  & 	2.0            &    $0.27 \times 0.17$    \\
	    \hline
		DK Tau   & \multirow{3}{*}{}continuum  & 	2.7e-02                &   $ 0.28 \times 0.14 $  \\
	                                        &$^{12}$CO  & 	4.4                 &    $0.24 \times 0.15$    \\
	                                        &$^{13}$CO  & 	2.5                 &    $0.26 \times 0.15$   \\
	                                        &C$^{18}$O  & 	2.1                 &    $0.26 \times 0.15$    \\
	    \hline
		GK Tau   & \multirow{3}{*}{}continuum  & 	3.4e-02                &    $0.24 \times 0.15$   \\
	                                        &$^{12}$CO  & 	12.2                 &   $0.26 \times 0.16$    \\
	                                        &$^{13}$CO  & 	5.1             &    $0.27 \times 0.17$    \\
	                                        &C$^{18}$O  & 	4.1                 &    $0.27 \times 0.17$    \\
	    \hline
		HK Tau   & \multirow{3}{*}{}continuum  & 	2.8e-02                &    $0.22 \times 0.14$    \\
	                                        &$^{12}$CO  & 4.7                 &    $0.24 \times 0.15$  \\
	                                        &$^{13}$CO  & 	3.7                &    $0.25 \times 0.15$   \\
	                                        &C$^{18}$O  & 	1.7                 &    $0.25 \times 0.15$    \\
	    \hline
		HN Tau   & \multirow{3}{*}{}continuum  & 2.7e-02               &    $ 0.19 \times 0.12 $  \\
	                                        &$^{12}$CO  & 6.9                 &   $ 0.19 \times 0.12 $  \\
	                                        &$^{13}$CO  & 	3.3                &   $ 0.20 \times 0.13 $    \\
	                                        &C$^{18}$O  & 	2.4                 &   $ 0.20 \times 0.13 $   \\
	    \hline
		T Tau    & \multirow{3}{*}{}continuum  & 4.3e-02              &   $0.13 \times 0.12 $  \\
	                                        &$^{12}$CO  &  41                 &    $0.14 \times 0.14$    \\
	                                        &$^{13}$CO  & 	4.6              &    $0.15 \times 0.14 $  \\
	                                        &C$^{18}$O  & 3.7                 &   $0.15 \times 0.14 $    \\
	    \hline
		UY Aur   & \multirow{3}{*}{}continuum  & 2.7e-02              &    $0.24  \times 0.13 $   \\
	                                        &$^{12}$CO  & 14                 &  $ 0.26 \times 0.15 $   \\
	                                        &$^{13}$CO  & 	4.5               &   $0.28 \times 0.15 $\\
	                                        &C$^{18}$O  &  3.7                &   $ 0.27 \times 0.15$   \\
	    \hline
		UZ Tau   & \multirow{3}{*}{}continuum  & 2.8e-02               &   $0.23 \times 0.14$  \\
	                                        &$^{12}$CO  & 4.2                 &   $0.24 \times 0.15 $ \\
	                                        &$^{13}$CO  & 4.6               & $0.25 \times 0.15$   \\
	                                        &C$^{18}$O  & 2.4                &    $0.25 \times 0.15$    \\
	    \hline
		V710 Tau & \multirow{3}{*}{}continuum  & 3.4e-02               &    $0.27 \times 0.23$  \\
	                                        &$^{12}$CO  & 5.6                 &   $0.28 \times 0.25 $   \\
	                                        &$^{13}$CO  & 4.5               &    $0.29 \times 0.26 $ \\
	                                        &C$^{18}$O  & 3.1                &   $ 0.30 \times 0.26$   \\
	    
		\hline
	\end{tabular}
\label{tab:rms_beam_cont}
\end{table}

In this work, in order to estimate disc radii and test tidal truncation models, we analyse the $^{12}$CO emission. In Table \ref{tab:rms_beam_cont} we report the rms of the resulting continuum image and of the zeroth moment image of the CO isotopologues, and the beam sizes.
Figure \ref{fig:mysampleMOM0} shows the $^{12}$CO zeroth moment images of the 10 systems observed in our program, of which three (GK~Tau, UY~Aur, and T~Tau) are excluded from the analysis provided in this paper (see Sect.~\ref{sect:sample}). The figure shows 5 times the rms of the continuum emission with white contours.

We detect and spatially resolve discs around all primary stars both in continuum and $^{12}$CO emission. 
For discs around the secondary stars, three discs (V710 Tau B, HN Tau B, DH Tau B) are not detected with neither dust continuum nor CO emission, CIDA 9 B disc does not show any CO emission despite hosting a dusty disc detected in the continuum, and both components are detected in the remaining sample. 
Table \ref{tab:detections} shows for each systems which components are detected in the continuum emission and in CO emission.

\begin{table} 
\centering 
\small
\caption{Detections of continuum and CO isotopologues.}
\begin{tabular}{l*{5}{c}}

 &   Cont det & $^{12}$CO det & $^{13}$CO det & C$^{18}$O det  \\
 
\hline\hline
CIDA 9 & A, B$^{*}$ & A & A & A \\
DH Tau & A & A & A & A$^{*}$  \\
DK Tau & A, B & A, B & A$^{\dagger}$ & -\\
GK Tau & A & A & A & - \\
HK Tau  & A, B & A, B & A, B & - \\
HN Tau & A & A & - & -  \\
T Tau & N, Sa, Sb &  N, Sa, Sb & N, Sa, Sb & N, Sa, Sb \\
UY Aur & A, B & A, B   & A, B & -\\
UZ Tau & E, Wa$^{*}$, Wb$^{*}$ & E, Wa, Wb & E, Wa, Wb & E$^{*}$, Wa$^{*}$ \\
V710 Tau & A & A & A & A  \\
\hline
\end{tabular}
\tablefoot{For each system we indicate the detected components. Note that GK Tau is a wide binary system with GI Tau: the label `A' refers to the disc around GK Tau. $^{*}$Not resolved; $^{\dagger}$ marginally resolved.}\label{tab:detections}
\end{table}

\begin{figure*}[]
    	\centering
	    \includegraphics[width=\textwidth,keepaspectratio]{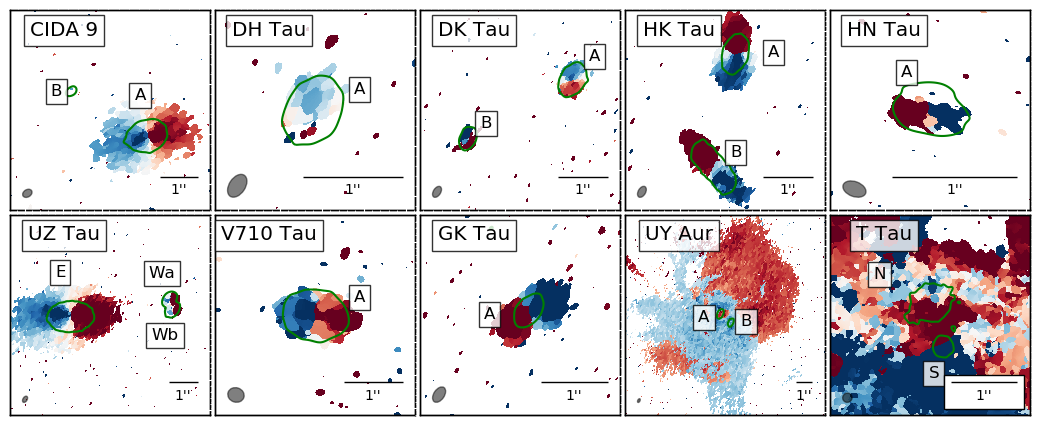}
	\caption{$^{12}$CO velocity map images of the discs created with \textit{bettermoments} quadratic method \citep{2018Teague_Foreman-Mackey}. A 4 sigma clipping process was applied to all images.
	The FWHM beam size is shown in the bottom left of each panel. Each image is scaled so that the maximum and the minimum are equal to the systemic velocity of the primary target $\pm2$ km/s. Green contours show 5 times the rms of the continuum emission. The components of the systems are labeled. The label for any undetected secondary components in the continuum emission is not shown. }\label{fig:mysampleQUADRATIC}
\end{figure*}

Figure \ref{fig:mysampleQUADRATIC} shows the velocity maps of the observed targets, including the three systems not analysed in this work. In particular, the ``quadratic'' method to image the velocity maps, created with \textit{bettermoments} tools \citep{2018Teague_Foreman-Mackey} and applying a 4 sigma clipping process, is shown.
Excluding T Tau and UY Aur systems, a pattern compatible with Keplerian rotation is observed in all detected circumprimary and circumsecondary discs, a few of which are already well known from the literature (e.g. HK Tau discs, \citealp{2014Jensen_Akeson}).
The spectra and the first moment images of all targets in the sample are presented in Appendix~\ref{sec:spectra+maps}.

\begin{figure}[]
    	\centering
	    \includegraphics[width=0.45\textwidth]{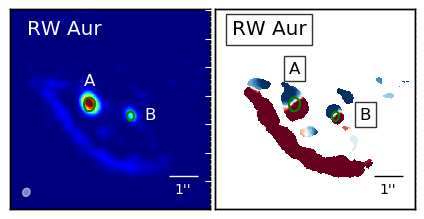}
	\caption{Maps of the RW Aur system observed by \cite{2018Rodriguez+}. Left: $^{12}$CO zeroth moment. The color scale has the peak flux as the maximum, and two times the image rms as minimum. Right: $^{12}$CO first velocity maps created with \textit{bettermoments} quadratic method \citep{2018Teague_Foreman-Mackey} and with a 4 sigma clipping process. Image scaled so that the maximum and the minimum are equal to the systemic velocity of the primary target $\pm2$ km/s.}\label{fig:mysampleRWAur}
\end{figure}

To this sample, we add the high angular resolution observation of the tidally disrupted protoplanetary discs of the RW Aurigae system analysed by \cite{2018Rodriguez+}.
As part of ALMA Cycle 3 and 4 projects 2015.1.01506.S  (PI Rodriguez) and 2016.1.00877.S (PI Rodriguez), RW Aur was observed in Band 6 (225 GHz) with the 12m array for a total integration time of 311.17 min and with the 7m array for a total additional integration time of 217.23 min. Additionally, RW Aur was observed in Band 7 (338 GHz) for a total integration time of 22.8 min.
The data were calibrated by the NAASC, and two rounds of phase-only self-calibration were applied to each set of observations. The $^{12}$CO 2-1 and 3-2 lines were imaged at a velocity resolution of 0.5 km/s, using a Briggs weighting with a robust value of 0.5, and with a \textit{uv}-taper of $0\farcs6 \times 0\farcs25$, resulting in a synthesised beam of $0\farcs30 \times 0\farcs25$ and an rms of 1.5 mJy/beam.
Figure~\ref{fig:mysampleRWAur} shows the $^{12}$CO zeroth moment image (left), the $^{12}$CO first moment map (central) and the $^{12}$CO velocity map imaged with \textit{bettermoments} quadratic method (right) for RW Aur system.

The channel maps and the spectra of the $^{12}$CO emission for all targets in the sample are reported in Appendix~\ref{sec:spectra+maps} and are useful to check for absorption due to foreground cloud material in the data. Signatures of partial absorption around the systemic velocity is observed in almost all discs; absorption in the red-shifted velocities is observed in the disc around CIDA 9 A, while the red-shifted velocities around DH Tau A are totally absorbed by the cloud below the continuum level. These absorption features are in line with previous observations \citep[e.g.,][]{2013guilloteau,2014Akeson_Jensen, 2019czekala}. Table~\ref{tab:cloudAbs} summarizes the absorbed velocities in the disc $^{12}$CO emission.

\begin{table} 
\centering 
\small
\caption{Systemic velocities (in km/s; second column) and absorbed velocities (in km/s; third and fourth columns) in the disc $^{12}$CO emission.}
\begin{tabular}{l*{4}{c}}

 &   Systemic Velocity & Primary  & Secondary  \\
 
\hline\hline
CIDA 9 & $5.93^{+0.010}_{-0.02}$& -$^{*}$ $^{\ddag}$ & non resolved \\
DH Tau & $5.014^{+0.013}_{-0.013}$&5.4-$^\dagger$ &  non detected \\
DK Tau & $5.081^{+0.002}_{-0.003}$&-$^{\ddag}$ & 3.0-7.0  \\
HK Tau  & $5.34^{+0.12}_{-0.12}$&5.2-7.2 $^{*}$ & 5.5-6.9 \\
HN Tau & $5.320^{+0.004}_{-0.002}$&3.8-5.6 & non detected \\
RW Aur & $\sim 6.0$ &-$^{\ddag}$ & -$^{\ddag}$ \\
UZ Tau & $4.95^{+0.03}_{-0.03}$ &5.5-6.3 &  Wa,Wb: 5.4-7.2 \\
V710 Tau & $6.467^{+0.005}_{-0.001}$&5.8-6.9  &  non detected \\
\hline
\end{tabular}
\tablefoot{The systemic velocities shown in the first column are estimated from the Keplerian modeling of the circumprimary disc velocity maps (see Section~\ref{sec:eddy} and Table~\ref{tab:estimatedDiscProp} for further details). $^{*}$ Sign of partial absorption in the red-shifted velocities. $^\dagger$ Red-shifted velocities totally absorbed. $^{\ddag}$ Sign of partial absorption around the systemic velocity.}
\label{tab:cloudAbs}

\end{table}

\section{Data analysis}\label{sect:analysis}
In this Section, we  estimate the radii of the discs. In order to measure this quantity, one needs first to determine the position angle and inclination of the discs. 
Once these disc parameters are derived an analysis on the image plane is performed in order to calculate the disc radii.

\subsection{Geometrical properties of discs from Keplerian modeling}\label{sec:eddy}

The geometrical properties of interest for each disc are the disc inclination $i$, its position angle PA, and the target centre (ra,dec). 
To estimate these properties, we used the \textit{eddy} tool \citep{2019Teague}, that aims at reconstructing the rotational profile of a gas emission line by fitting a Keplerian rotation pattern to the velocity map of the observations, taking into account the relative uncertainties calculated with the \textit{bettermoments} tool \citep{2018Teague_Foreman-Mackey} and assuming a geometrically thin disc.

Since there is a degeneracy between the inclination $i$ and the stellar mass $M_\star$, the stellar mass was fixed to the values reported in Table \ref{tab:ch4paramDisc}. In this way, \textit{eddy} would fit five free parameters (ra, dec,
 PA, $i$ and the systemic velocity $v_\mathrm{LSR}$), given two fixed parameter ($M_\star$ and the distance $d$ of the disc from the observer).
To account for the impact of the uncertainty on \mstar\, on the estimate of the parameters of the discs, we performed for each target three fits, one adopting the assumed value of $M_\star$, and one each for the $\pm$1$\sigma$ values of $M_\star$ given the uncertainty (see Table \ref{tab:ch4paramDisc}). The fit parameters are initialised using moment 0 value, for the target center, and the values by \citet{2019Manara+}, for inclination and position angle. Flat priors for all parameters were chosen, with Gaussian priors only for the target center. The number of walkers and steps needed to achieve the convergence of the chain were typically $200$ walkers and $\sim1000$-$2000$ steps, with the last $\sim800$-$1000$ steps used to sample the posterior distribution.

We account for cloud absorption in discs excluding from the fit the velocities absorbed by the cloud (see Table~\ref{tab:cloudAbs}).
Since the disc around DH Tau A is heavily absorbed in the red-shifted velocities, we only fit the $^{13}$CO velocity map of DH~Tau.

We did not fit the disc around the secondary star of the HK Tau system. This disc is almost edge-on, so that the assumption of geometrically thin disc emission cannot be applied and the vertical height of the disc must be considered. Since the HK Tau system is well known in the literature, we assumed the position angle and the inclination estimated in a previous work by \cite{2011McCabe+} (see also \citealt{2014Jensen_Akeson}): PA=$(42 \pm 0)$\si{\degree} and $i = (85 \pm 1) \si{\degree}$. 

To avoid issues with the convergence of the \textit{eddy} fit, we estimated the source centre of the circumsecondary discs fitting two elliptical Gaussian components to the $^{12}$CO image with the CASA task \textit{imfit}. All other parameters of circumsecondary discs were estimated through \textit{eddy}.
Table~\ref{tab:estimatedDiscProp} reports the estimated disc properties for each disc. 

\begin{figure}[h]
    	\centering
    	\includegraphics[trim=0mm 0cm 22.5cm 0mm,clip,width=0.5\textwidth]{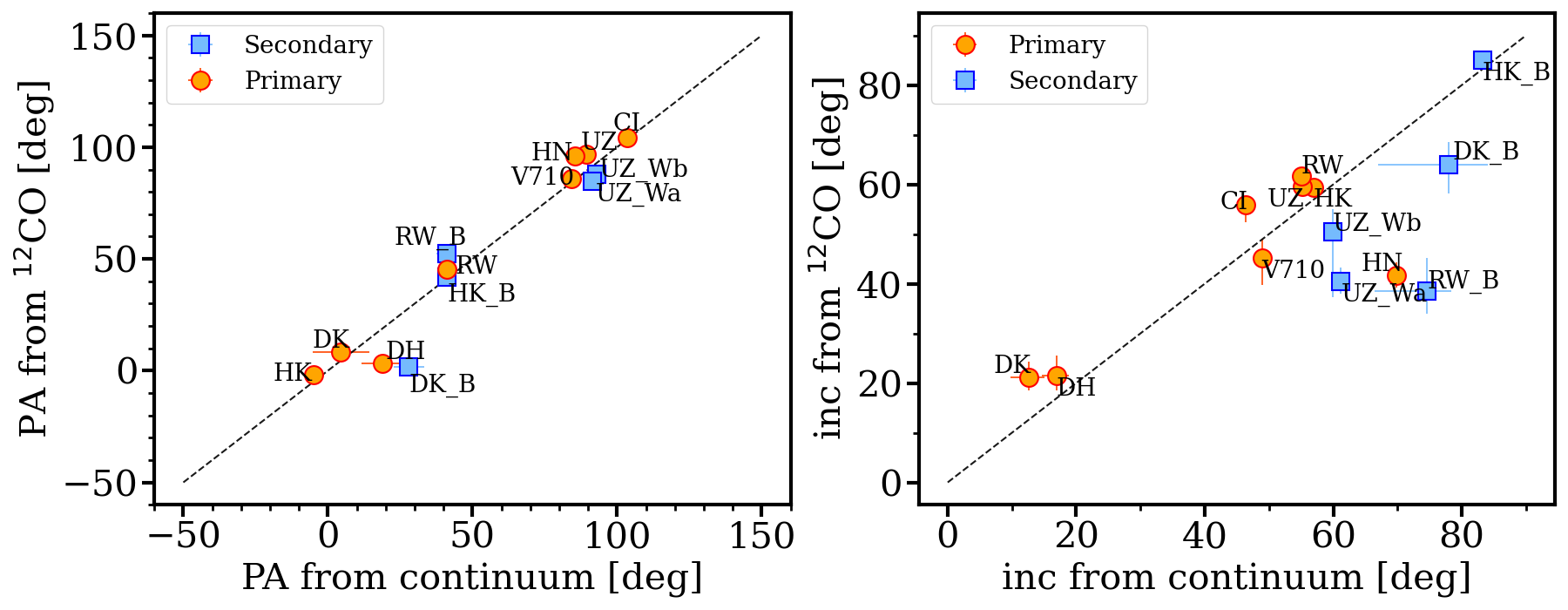}
	    \includegraphics[trim=23.2cm 0cm 0cm 0mm,clip,width=0.5\textwidth]{pa_inc.png}
	\caption{Position angle (top) and inclination (bottom) estimated through \textit{eddy} fitting on $^{12}$CO first moment maps compered with that one estimate by \cite{2019Manara+} through $(u,v)$-plane modelling of the continuum data with \textit{Galario}. Red circles refer to circumprimary discs, while blue squares refer to circumsecondary discs. }\label{fig:pa_inc}
    \end{figure}

The agreement between the source centres obtained here with those derived by \cite{2019Manara+} is good when accounting for the typical proper motion of the targets. 
We also compared the position angle and inclination obtained fitting the gas emission with \textit{eddy} with the values obtained by \citet{2019Manara+} fitting the continuum data in the uv-plane. This comparison is shown in Figure \ref{fig:pa_inc}. The values obtained with the two methods are in agreement within the (3$\sigma$) uncertainties, with the exception of the circumsecondary discs around DK Tau B, RW Aur B and UZ Tau Wa, where the estimates with both method are uncertain due to the small size of the targets.
Another exception, probably due to some issues in the \textit{Galario} fitting, is the circumprimary disc around HN Tau A.
The agreement within the uncertainties of the estimates with the two different methods confirms that the assumptions on the stellar masses are reasonable, except possibly in the four cases just mentioned. Indeed, \cite{2019Manara+} did not need any assumption on stellar masses to fit the continuum, whereas in the Keplerian modelling performed here a value of $M_\star$ was assumed.

\begin{table*} 
\centering 
\caption{Source centres (ra,dec), inclination $i$, position angle PA and systemic velocity $v_{LSR}$ for all discs in the analysed sample. }  
\begin{tabular}{l*{6}{c}}

 &   ra &     dec & inc [deg] & PA [deg] & $v_{lsr}$ [m/s]\\
 
\hline\hline
CIDA 9 A & 05h05m22.8260 & +25d31m30.5550 & $56.0^{+1.4}_{-3.5}$ & $284.2^{+0.7}_{-0.4}$ & $5930^{+10}_{-20}$\\
DH Tau A & 04h29m41.5610 & +26d32m57.7442 & $21.4^{+4.2}_{-3.0}$ & $183.3^{+1.0}_{-1.3}$ & $5014^{+13}_{-13}$ \\
DK Tau A & 04h30m44.2509 & +26d01m24.2994 & $21.1^{+3.3}_{-2.7}$ & $188.2^{+0.1}_{-0.1}$ & $5081^{+2}_{-3}$\\
DK Tau B & 04h30m44.4048 & +26d01m23.1589 & $64.0^{+4.6}_{-5.8}$ & $181.6^{+1.7}_{-2.6}$ & $6000^{+390}_{-240}$\\
HK Tau A & 04h31m50.5769 & +24d24m17.3400 & $59.48^{+0.6}_{-1.0}$ & $358.1^{+2.4}_{-4.5}$ & $5340^{+120}_{-120}$\\
HK Tau B & 04h31m50.6122 & +24d24m15.1045 &  $85^{+1}_{-1}$ & $42.0^{+0.0}_{-0.0}$ & $\sim 6000$ \\
HN Tau A & 04h33m39.3782 & +17d51m51.9698 & $41.6^{+2.7}_{-2.4}$ & $92.07^{+0.04}_{-0.07}$ & $5320_{-2}^{+4}$\\
RW Aur A & 05h07m49.5716 & +30d24m04.7380 & $61.7^{+0.5}_{-0.8}$ & $225.38^{+0.03}_{-0.11}$ & $\sim 6000$\\
RW Aur B & 05h07m49.4597 & +30d24m04.3309 & $38.6^{+4.7}_{-6.6}$ & $232.2^{+0.4}_{-0.4}$ & $\sim 5200$\\
UZ Tau E & 04h32m43.0742 & +25d52m30.6460 & $59.4^{+0.1}_{-0.1}$ & $276.96^{+0.05}_{-0.03}$ & $4950^{+30}_{-30}$ \\
UZ Tau Wa & 04h32m42.8236 & +25d52m31.1959 & $40.5^{+3.0}_{-2.5}$ & $264.7^{+0.2}_{-0.3}$ & $6571.3^{+0.6}_{-0.5}$\\
UZ Tau Wb & 04h32m42.8159 & +25d52m30.8347 & $50.4^{+4.7}_{-13.1}$ & $267.9^{+0.4}_{-0.4}$ & $6390^{+50}_{-30}$\\
V710 Tau A & 04h31m57.8081 & +18d21m37.6092 & $45.1^{+3.8}_{-5.5}$ & $266.049^{+0.70}_{-0.03}$ & $6467_{-1}^{+5}$\\
\hline
\end{tabular}\tablefoot{The position angle PA is defined as the angle to the redshifted disc major-axis counting from North to East. The target centres for circumsecondary discs and for RW Aur A disc were estimated through the CASA tool \textit{imfit}. The disc inclination and position angle of HK Tau B are those reported in \cite{2011McCabe+}. The systemic velocity of HK Tau B is estimated from the spectrum (see Figure \ref{fig:maps}), while $v_\mathrm{lsr}$ of RW Aur discs are assumed from \cite{2018Rodriguez+}. All other parameters were estimated through \textit{eddy} tool.}
\label{tab:estimatedDiscProp}
\end{table*}

\subsection{Disc radii estimate: the cumulative flux technique}\label{sec:radii_estim}

To estimate the disc radii, we perform an image plane analysis using the cumulative flux technique \citep[for example, see][]{2018Ansdell+}\footnote{For reference, this method has been called "curve of growth" by \cite{2018Ansdell+}.}, computing aperture photometry on the zeroth moment images of the $^{12}$CO emission and on the continuum images using the CASA tool \textit{imstat}. This technique is applied on the continuum emission (both on new data and on data from \citealp{2019Manara+}) and on the $^{12}$CO zeroth moment image of each target, both primary and secondary (tertiary), when detected.

The cumulative flux technique consists of calculating the flux of the source at increasingly larger radii. In particular, for each target the flux is calculated summing over concentric annuli centred in the source centre and corrected for the projected position of the disc (PA and inclination from Table~\ref{tab:estimatedDiscProp}). The annuli are increased in step of $\mathrm{d}r=$ 0\farcs05 (one-third of the angular resolution of the data), and the flux uncertainty in each annulus is estimated as the standard deviation of the fluxes in 100 random annuli selected well outside the disc.
We consider the error on the disc parameters calculating three cumulative flux curves assuming in each of these either the maximum or minimum allowed values for the position angle and inclination of the ellipses given the errors on these two quantities. In this way, both the uncertainty due to the (statistical) standard deviation of fluxes (random annuli in the field) and to the uncertainty on the stellar masses (uncertainty on PA and $i$) are considered.

Once the maximum flux is calculated, we estimate the disc radius $R_\mathrm{disc}$ (or $R_{95}$) as the radius containing $95\%$ of the total flux and the effective disc radius $R_\mathrm{eff}$ (or $R_{68}$) as that containing $68\%$ of the total flux. The uncertainty on the radius is calculated considering the statistical error on the maximum flux and, since the error in the radius estimate must be at least half of the resolution, we summed in quadrature this error with 0\farcs075.

A different procedure than the above described was applied in the case of the HK~Tau~B disc. For this target, known also from the literature to be a edge-on disc \citep[e.g.,][]{2011McCabe+, 2014Jensen_Akeson}, the vertical structure in the gas emission is more extended than the continuum. 
Assuming an inclination of $\sim 85$\si{\degree} \citep[][]{2011McCabe+, 2014Jensen_Akeson} would lead to an estimated gas radius much larger than the separation between the components of the HK Tau system.
To correct for this effect, the disc radius is measured along the major axis, centering on the source rectangles with increasing width and fixed height (i.e. fixed to the HK Tau beam size in that direction -- $\sim 0\farcs25$), and rotated by the position angle of the disc. No significant change in the estimated disc radius was found modifying the height of the rectangles from $0\farcs10$ to  $0\farcs25$.
The dust disc size is instead affected by optical depth effects. Being an almost edge-on disc, the brightness profile of HK Tau B is less peaked at small radii (near the disc centre) because of dust self-absorption \citep{2020Villenave+}. The observed profile is dominated by the emission from larger radii, where the disc is colder, and thus resulting in a larger estimate of the dust disc radius (see Section~\ref{sec:dust-gas} for further details).

		Since RW Aur system is known to experience interaction between the components \citep{2018Rodriguez+}, to avoid considering the emission due to the interaction, the $^{12}$CO total flux of the RW Aur A and B discs was assumed to be the enclosed flux closest to the one estimated through the CASA tool \textit{imfit} (two-Gaussian fit) -- $3.661 \pm 0.094$ Jy km/s and $0.949 \pm 0.076$ Jy km/s, respectively.
		
	    Since the southern part of the $^{12}$CO emission of DH Tau circumprimary disc is absorbed by the cloud (Figure \ref{fig:maps}), we applied the cumulative flux technique only on the northern part, masking the emission from the southern half of the disc. Since the position angle of DH Tau A disc is $\sim 180 \si{\degree}$, we masked the emission observed below a line with PA$=90 \si{\degree}$ centred on the source centre.
	    We also masked the southern part of the UZ Tau Wa disc and the northern part of the UZ Tau Wb disc, both in the continuum and $^{12}$CO emission, to exclude emission arising from the other component of the system.
	    Since the position angle of UZ Tau Wa and Wb discs is $\sim 270 \si{\degree}$, we drew two lines with PA$=90 \si{\degree}$ centred on UZ Tau Wa disc centre and UZ Tau Wb disc centre and we masked the emission observed below and above each line, respectively.

\begin{figure*}[]
    	\centering
	    \includegraphics[trim=0mm 0.5cm 25cm 0mm,clip,width=0.95\textwidth]{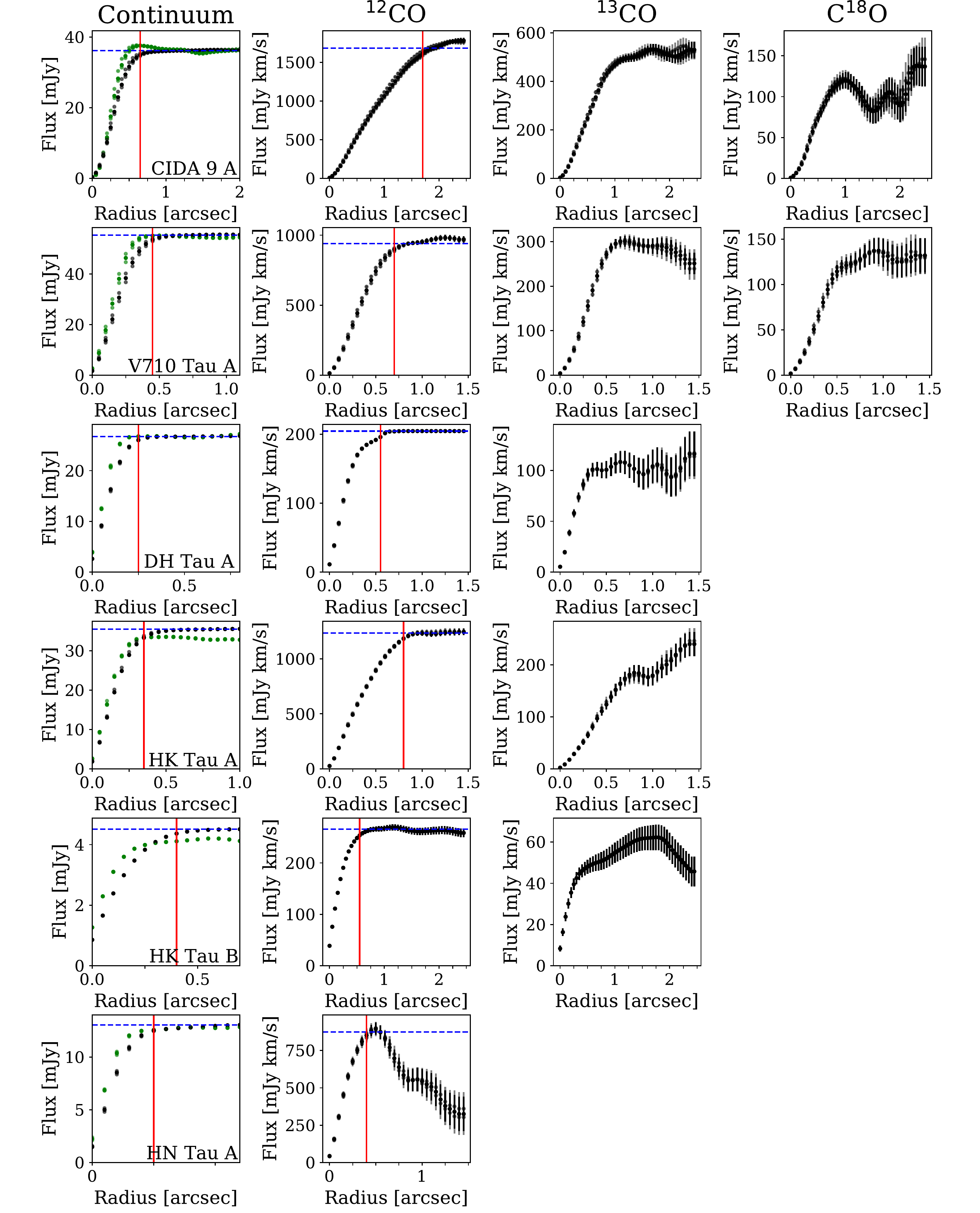}
	\caption{Cumulative fluxes as a function of aperture radius estimated using the cumulative flux technique. Each panel shows the fluxes for a different data-set/image (the name of the target for each row is shown in the lower right corner of the first column). The first column shows the continuum emission of the data from \cite{2019Manara+} (green dots) and the continuum emission from the new data (black dots). The second, third and fourth columns show the $^{12}$CO, $^{13}$CO and C$^{18}$O moment 0 emission from the new data, respectively. The blue dashed lines show the estimated disc total fluxes; the red solid lines show the disc radii, defined as radii containing the $95\%$ of the total flux (see values reported in Table~\ref{tab:fluxes} and \ref{tab:radii}). In the first column, only the total fluxes and disc radii estimated from the new continuum data are shown.}\label{fig:cog}
\end{figure*}

\begin{figure*}[]
    	\centering
	    \includegraphics[trim=1cm 3.cm 7cm 9.cm,clip,width=0.93\textwidth]{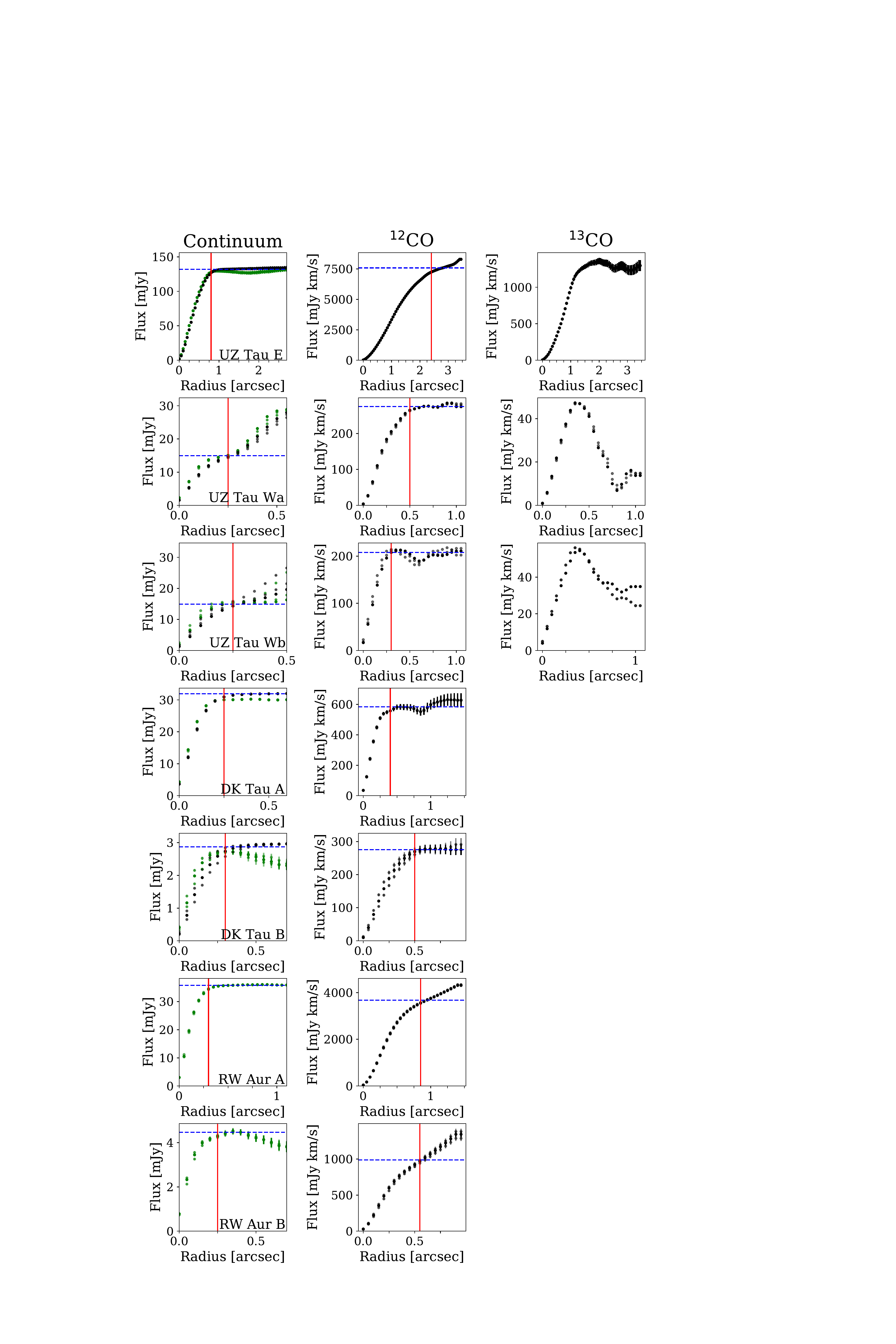}
	\caption{Cumulative flux curves as a function of aperture radius estimated using the cumulative flux technique. Same as in Figure~\ref{fig:cog} for the remaining discs in the sample. The total fluxes and disc radii estimated from the continuum data from \citep{2019Manara+} are shown in the case of RW Aur A and B discs (first column, last two rows). In all the other cases, the total fluxes and disc radii estimated from new continuum data are shown.}\label{fig:cog2}
\end{figure*}

	    Figure \ref{fig:cog} and \ref{fig:cog2} show the cumulative flux curves for each disc in the sample. The first columns show the cumulative sum of the fluxes measured on the continuum emission from the data by \cite{2019Manara+} and the new data; the second, third and fourth columns show the cumulative flux relative to the $^{12}$CO, $^{13}$CO, and C$^{18}$O emission, respectively.
	    The blue lines show the total fluxes (see Table~\ref{tab:fluxes}), while the red lines show the disc radii ($R_{95}$, see Table~\ref{tab:radii}).
	    

\begin{figure*}[h]
    	\centering
	    \includegraphics[width=0.85\textwidth]{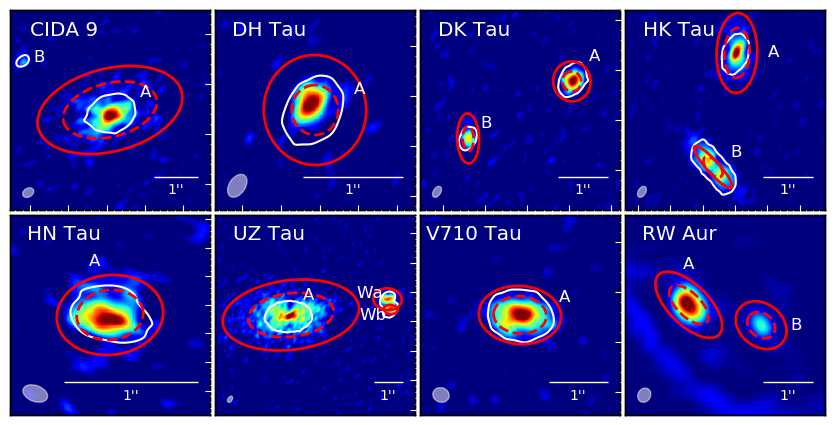}
	\caption{$^{12}$CO zeroth moment images of the discs with overlapped ellipses including the $95\%$ (solid) and $68\%$ (dashed) of the total flux. The ellipses are centred on the target source centre. All bars in the bottom right of each panel are 1\farcs0 long, which is $\sim 140$ au at the distance of Taurus. The beam FWHM size is shown in the bottom left of each panel. In each image, the color scale has the peak flux as the maximum, and two times the image rms as minimum. White contours show 5 times the rms of the continuum emission. }\label{fig:mom0_radii}
\end{figure*}

Table \ref{tab:radii} shows the estimated radii and their error for each disc, while Figure \ref{fig:mom0_radii} shows the ellipses including the $95\%$ (solid) and $68\%$ (dashed) of the total flux overlapped to the zeroth moment maps.
		As shown by the red solid ellipses, the radii estimated with the cumulative flux technique get the really outer extent of the discs, where little emission is detected, giving the possibility to test tidal truncation models.
		Table~\ref{tab:fluxes} shows the total fluxes we estimated for each discs with the cumulative flux technique applied on the continuum image and on the $^{12}$CO, $^{13}$CO and C$^{18}$O zeroth moment images.
		
		The effective disc radii estimated for the RW Aur system are in very good agreement with the ones reported in \cite{2018Rodriguez+}, estimated using CASA built-in measurement tool \citep{2007McMullin+}, both for the circumprimary and the circumsecondary discs $-$ $\sim 58$ au and $\sim 38$, respectively, at a distance of 140 pc (assumed by \citealt{2018Rodriguez+}).

\begin{table*}[h!]
    \centering
     \caption{Radii estimated with the cumulative flux technique}
    \begin{tabular}{ccccc}
        \multirow{2}{*}{Target}  & & \multirow{2}{*}{$ ^{12}$CO [\arcsec]} & \multirow{2}{*}{Continuum [\arcsec]}  & Continuum from  \\
        & &   &   & \cite{2019Manara+} [\arcsec] \\
       \hline \hline
         \multirow{2}{*}{CIDA 9 A}& $R_{68}$ & $1.10^{+0.08}_{-0.09}$  & $0.40^{+0.08}_{-0.08}$ & $0.35^{+0.08}_{-0.08}$ \\ & $R_{95}$ &   $1.70_{-0.09}^{+0.13}$ &  $0.650^{+0.08}_{-0.08}$ &  $0.50^{+0.08}_{-0.08}$ \\
         
         \hline
         
        \multirow{2}{*}{DH Tau A}& $R_{68}$ & $0.25^{+0.08}_{-0.08}$ & $0.15^{+0.08}_{-0.08}$ &  $0.10^{+0.08}_{-0.08}$ \\ & $R_{95}$ & $0.55_{-0.08}^{+0.08}$ & $0.25_{-0.08}^{+0.08}$ & $0.20_{-0.08}^{+0.08}$ \\
       \hline
        \multirow{2}{*}{DK Tau A}& $R_{68}$ & $0.20_{-0.08}^{+0.08}$ & $0.15_{-0.08}^{+0.08}$ & $0.10_{-0.08}^{+0.08}$ \\ & $R_{95}$ & $0.40^{+0.13}_{-0.13}$ & $0.25^{+0.08}_{-0.08}$ & $0.20^{+0.08}_{-0.08}$ \\
        \hline
    
        \multirow{2}{*}{DK Tau B}& $R_{68}$  & $0.25_{-0.08}^{+0.09}$ & $0.20_{-0.09}^{+0.08}$ & $0.10_{-0.08}^{+0.08}$ \\ & $R_{95}$ & $0.50_{-0.13}^{+0.09}$ & $0.30_{-0.08}^{+0.09}$ & $0.20_{-0.08}^{+0.09}$ \\ 
        \hline
        
        \multirow{2}{*}{HK Tau A}& $R_{68}$ & $0.50_{-0.09}^{+0.08}$ & $0.20_{-0.08}^{+0.08}$ & $0.15_{-0.08}^{+0.08}$ \\ & $R_{95}$ & $0.80_{-0.09}^{+0.09}$ & $0.35_{-0.08}^{+0.08}$ & $0.30_{-0.09}^{+0.08}$ \\
        \hline
        
        \multirow{2}{*}{HK Tau B}& $R_{68}$  & $0.25_{-0.08}^{+0.08}$ & $0.20_{-0.08}^{+0.08}$ & $0.10_{-0.08}^{+0.08}$ \\ & $R_{95}$ & $0.55_{-0.09}^{+0.13}$ & $0.40_{-0.08}^{+0.08}$ & $0.20_{-0.08}^{+0.09}$ \\ 
        \hline
        \multirow{2}{*}{HN Tau A}& $R_{68}$ & $0.25_{-0.09}^{+0.08}$ & $0.15_{-0.08}^{+0.08}$ & $0.10_{-0.08}^{+0.08}$ \\ & $R_{95}$ & $0.40_{-0.09}^{+0.09}$ & $0.25_{-0.08}^{+0.08}$ & $0.20_{-0.08}^{+0.08}$ \\
\hline
        
        \multirow{2}{*}{RW Aur A}& $R_{68}$  & $0.50_{-0.09}^{+0.08}$ & - & $0.15_{-0.08}^{+0.08}$ \\ & $R_{95}$ & $0.85_{-0.09}^{+0.13}$ & - & $0.30_{-0.08}^{+0.08}$ \\
        \hline
        
        \multirow{2}{*}{RW Aur B}& $R_{68}$  & $0.30_{-0.08}^{+0.09}$ & - & $0.10_{-0.08}^{+0.08}$ \\ & $R_{95}$ & $0.55_{-0.13}^{+0.13}$ & - & $0.25_{-0.09}^{+0.09}$ \\
        \hline
        
         \multirow{2}{*}{UZ Tau E}& $R_{68}$  & $1.50_{-0.09}^{+0.08}$  & $0.50_{-0.08}^{+0.08}$  & $0.45_{-0.08}^{+0.08}$ \\ & $R_{95}$ & $2.40_{-0.13}^{+0.13}$ & $0.80_{-0.09}^{+0.09}$ & $0.70_{-0.08}^{+0.09}$ \\
        
        \hline
        
        \multirow{2}{*}{UZ Tau Wa}& $R_{68}$  & $0.30_{-0.08}^{+0.08}$ & $0.15_{-0.08}^{+0.08}$ & $0.10_{-0.08}^{+0.08}$ \\ & $R_{95}$ & $0.50_{-0.08}^{+0.09}$ & $0.25_{-0.08}^{+0.08}$ & $0.20_{-0.08}^{+0.08}$ \\ 
        \hline
        \multirow{2}{*}{UZ Tau Wb}& $R_{68}$  & $0.20_{-0.08}^{+0.08}$ & $0.10_{-0.08}^{+0.08}$ & $0.10_{-0.08}^{+0.08}$ \\ & $R_{95}$ & $0.30_{-0.08}^{+0.08}$ & $0.20_{-0.08}^{+0.09}$ & $0.15_{-0.08}^{+0.08}$ \\
        \hline
        \multirow{2}{*}{V710 Tau A}& $R_{68}$  & $0.45_{-0.08}^{+0.08}$ & $0.25_{-0.08}^{+0.08}$ & $0.20_{-0.08}^{+0.08}$ \\ & $R_{95}$ & $0.70_{-0.08}^{+0.09}$ & $0.45_{-0.08}^{+0.08}$ & $0.35_{-0.08}^{+0.08}$ \\
       
        \hline
        \hline
    \end{tabular}
   \tablefoot{We report the effective disc radius $R_{68}$ and the disc radius $R_{95}$ calculated applying the cumulative flux technique on the $^{12}$CO and continuum emission from new observations, and on the continuum emission data by \cite{2019Manara+}.}
    \label{tab:radii}
\end{table*}

\begin{table}
\small
\centering 
\caption{Total fluxes measured with the cumulative flux technique.}
\begin{tabular}{l*{5}{c}}

 Target & $F_{\rm cont}$ & $F(^{12} \rm CO)$&     $F(^{13} \rm CO)$  & $F(\rm C^{18} O)$ \\

\hline\hline

CIDA 9 A & $36.17^{+0.11}_{-0.10}$ & $1685.1^{+55.9}_{-39.8}$ &  $491.4^{+20.6}_{-17.4}$ &  $115.0^{+12.7}_{-10.0}$ \\
DH Tau A  & $26.71_{+0.06}^{-  0.07}$ &$409.02^{+0.06}_{-0.07}$ &  $101.3^{+7.3}_{-7.2}$ &  - \\
DK Tau A &  $31.94^{+0.07}_{-0.07}$ & $583.7^{+19.4}_{-19.5}$ & - & - \\
DK Tau B & $2.87^{+0.07}_{-0.10}$ &  $275.1^{+12.9}_{-15.5}$ & -  & - \\
HK Tau A & $35.50^{+0.08}_{-0.08}$ & $1233.5^{+31.6}_{-25.5}$ &  $181.8^{+18.0}_{-18.4}$ & - \\
HK Tau B & $4.51^{+0.02}_{-0.02}$ & $265.8^{+5.3}_{-5.3}$ & $48.8^{+3.6}_{-3.6}$ & - \\
HN Tau A & $13.06^{+ 0.08}_{-0.09}$ & $873.4^{+45.7}_{-50.8}$ &  - & - \\
RW Aur A & - & $3678^{+159}_{-54}$ &  - &  - \\
RW Aur B & - & $987.8^{100.1}_{-68.2}$ & - & - \\
UZ Tau E & $132.1^{+1.7}_{-1.7}$ & $7584^{+100}_{-101}$ &  $1333^{+38}_{-38}$ &  - \\
UZ Tau Wa &  $14.9^{+5.3}_{-3.6}$ & $550.6^{+1.3}_{-1.9}$ &  $94.2^{+1.4}_{-2.3}$ &  - \\
UZ Tau Wb &  $12.9^{+1.6}_{-0.2}$ & $415.5^{+8.5}_{-8.5}$ & $112.0^{+2.6}_{-2.6}$ &  - \\
V710 Tau A  &  $55.33^{+0.10}_{-0.09}$ & $939.5^{+16.3}_{-15.8}$ & $300.8^{+15.0}_{-18.2} $ &  $119.1^{+12.3}_{-11.0} $ \\
\hline
\end{tabular}
\tablefoot{Continuum fluxes are reported in mJy, line fluxes in mJy km/s.}\label{tab:fluxes}
\end{table}

\begin{figure*}[]
    	\centering
	    \includegraphics[width=\textwidth]{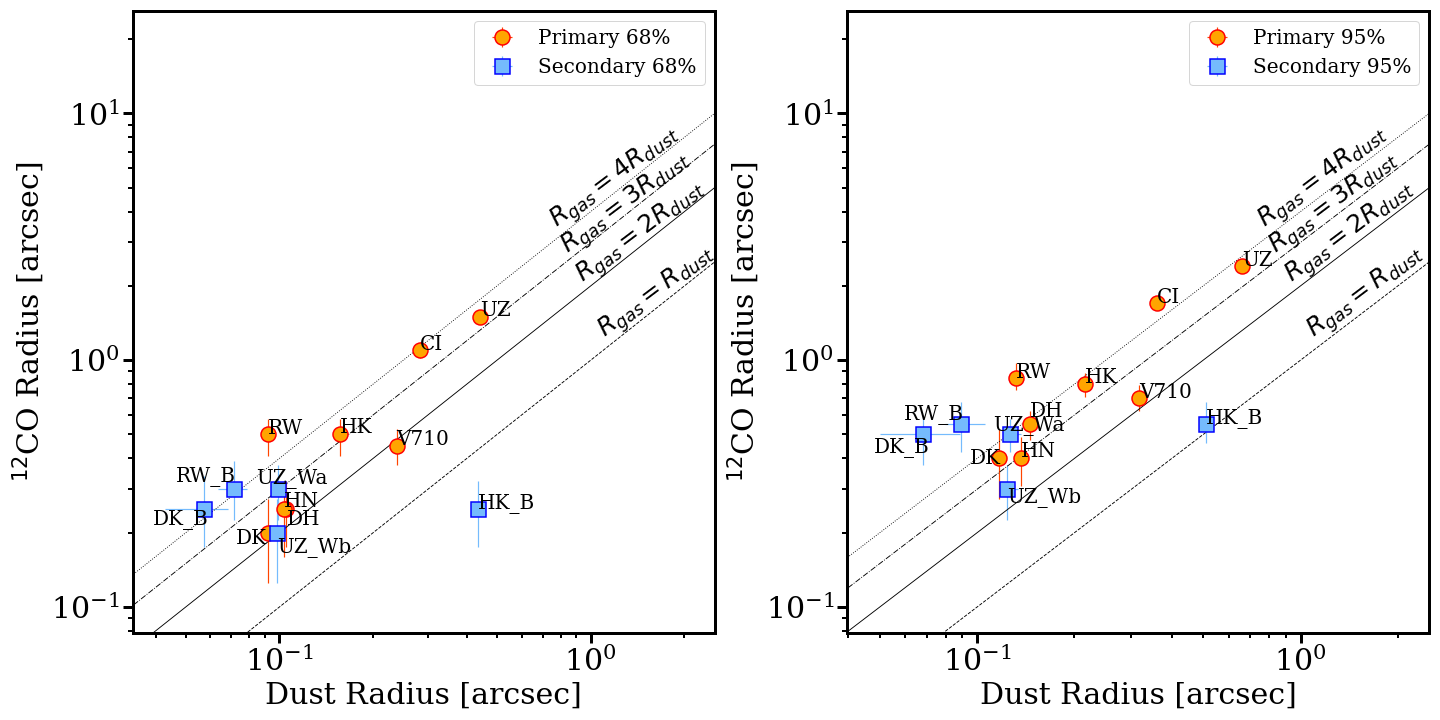}
	\caption{$^{12}$CO estimated radii as a function of dust radii obtained with \textit{(u,v)-}plane modelling of the continuum data by \cite{2019Manara+}. Dust and gas radii estimated as the radii including the 68\% (right) and the 95\% (left) of the total flux. The red circles are for circumprimary discs, while blue squares for the circumsecondary. }\label{fig:cfr1NewOld}
\end{figure*}

\begin{figure}[]
    	\centering
	    \includegraphics[width=0.5\textwidth]{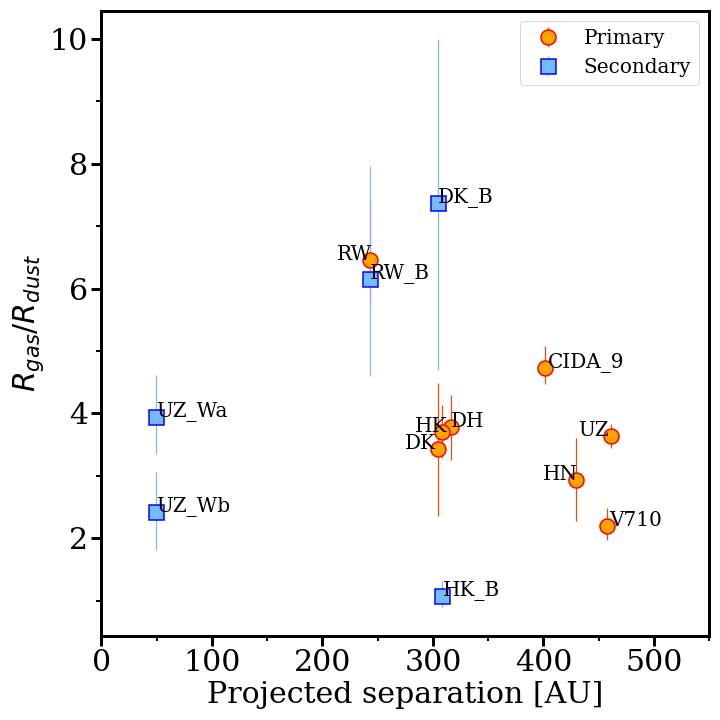}
	\caption{Gas-dust size ratios ($R_{\mathrm{95,gas}}/R_{\mathrm{95,dust}}$) as a function of the projected separation. The red circles are for circumprimary discs, while blue squares for the circumsecondary.}\label{fig:ratioVSsep}
\end{figure}

\section{Discussion}\label{sect:discussion}

\subsection{Comparing dust and gas disc radii}\label{sec:dust-gas}

The emission arising from the dusty component of discs, and thus the dust disc radius measurement, is regulated by several processes, like growth, fragmentation, and transport of dust. On the contrary, the gas emission extent is regulated by chemical processes and by the dynamics of the disc, thus in our case by the tidal interaction of the components of the multiple systems we are studying. 
Here we investigate whether the ratio between the size of the discs (defined as the radius containing the 95\% of the total flux) measured in the $^{12}$CO gas component and in the dust component (\rgas/\rdust) is different in systems where tidal truncation is at work with respect to more isolated systems. 

The ratio \rgas/\rdust \, has been studied in a limited number of targets in Taurus  \citep{2021Kurtovic}, and in a larger sample in the Lupus star-forming region, firstly analysed by \citet{2018Ansdell+}, and recently revised by \citet{2021Sanchis+}. The ratio \rgas/\rdust ~ in mostly isolated or wide binary systems in the Lupus star-forming region has a median and a mean value of $\sim$2.5 and $\sim$2.8 respectively, both when considering the radii at 68\% or 95\% of the total flux. In general, the larger gas sizes can be explained as a combination of the effect of radial drift of dust, and of the different opacity between the two components \citep[e.g.,][]{facchini17,trapman20}. \cite{2021Sanchis+} noted that the gas-dust size ratios and the stellar masses are uncorrelated, contrary to expectations from theoretical and observational works suggesting that the radial drift is more effective in discs around low-mass stars \citep[e.g.,][]{2016Pascucci+,2013Pinilla+}. For the sake of this work, the lack of any correlation between these two quantities assures that we can carry out a comparison between our results with those by \citet{2021Sanchis+} with no biases due to different stellar mass distributions. 

For the following discussion, we use the values of $R_\mathrm{dust}$ obtained by \cite{2019Manara+} by modelling the continuum data in the visibility plane using the \textit{Galario} library \citep{2018Tazzari+} because this makes the results directly comparable with \cite{2021Sanchis+}, who performed a similar analysis.

Our sample of multiple system reveals that the gas disc radii are typically $\sim2-4$ times larger than the dust disc radii, for both the effective gas-to-dust radii $R_{68, \mathrm{gas}}/R_{68, \mathrm{dust}}$ and the gas-to-dust disc radii $R_{95, \mathrm{gas}}/R_{95, \mathrm{dust}}$ (Figure \ref{fig:cfr1NewOld}). As shown by Figure~\ref{fig:cfr1NewOld}, HK Tau B disc is the biggest outlier in the gas-to-dust size ratio distribution, with $R_{95,\text{gas}}/R_{95,\text{dust}}\sim 1$. Since it is an almost edge-on disc, this small ratio may be due to optical depth effects, resulting in an overestimate of the dust radius and, as a consequence, in an underestimate of the gas-to-dust disc radius. We thus decided to exclude HK Tau B disc from the statistical analysis discussed below.

\begin{figure}
    
    	\centering
    	\includegraphics[width=0.45\textwidth]{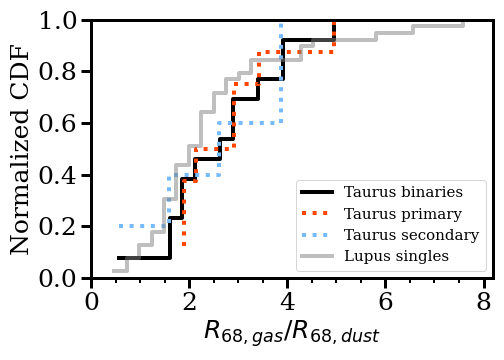}
    	\\
	    \includegraphics[width=0.45\textwidth]{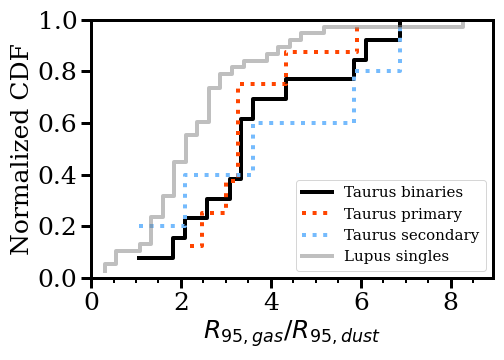}

	\caption{Normalized cumulative distribution function for the gas-to-dust effective disc sizes $R_{68, \mathrm{gas}}/R_{68, \mathrm{dust}}$ (top) and for the gas-to-dust disc sizes $R_{95, \mathrm{gas}}/R_{95, \mathrm{dust}}$ (bottom) for discs around single stars in Lupus (grey, \citealp{2021Sanchis+}) and discs around multiple stellar system in Taurus (black). Discs around primary stars in Taurus are shown in red, while circumsecondary discs in Taurus are shown in blue. }
	
	\label{fig:cdfratio}
\end{figure}

The average value (and relative standard deviation) of $R_{68, \mathrm{gas}}/R_{68, \mathrm{dust}}$ is $3.2\pm1.1$ ($3.1\pm1.1$ and $3.4\pm0.9$ for circumprimary and circumsecondary discs, respectively), and $R_{95, \mathrm{gas}}/R_{95, \mathrm{dust}}=4.2\pm1.6$ (with average values of $3.9\pm1.2$ for circumprimary discs and $5.0\pm1.9$ for the circumsecondary). The Kolmogorov-Smirnov (K-S) two-sided test for the null hypothesis that the samples of gas-to-dust size ratios around primary stars and around secondary stars are drawn from the same continuous distribution leads to a p-value $\sim0.06$ (for both the 68\% and the 95\% ratios), thus confirming that the circumprimary and circumsecondary discs are likely to be drawn from the same distribution (see Figure~\ref{fig:cdfratio}).

Figure \ref{fig:ratioVSsep} shows the ratio \rgas/\rdust\, as a function of the projected separation of the components in the systems. No evidence of a correlation between the size ratio and the projected separation is shown, suggesting that the gas-dust size ratio does not depend on the distance between the components in the system. Other features, such as the eccentricity of the orbits, need to be considered to explain the observed distribution of size ratios (see Section \ref{sec:ecc}).

\begin{figure*}[]
    	\centering
	    \includegraphics[width=\textwidth]{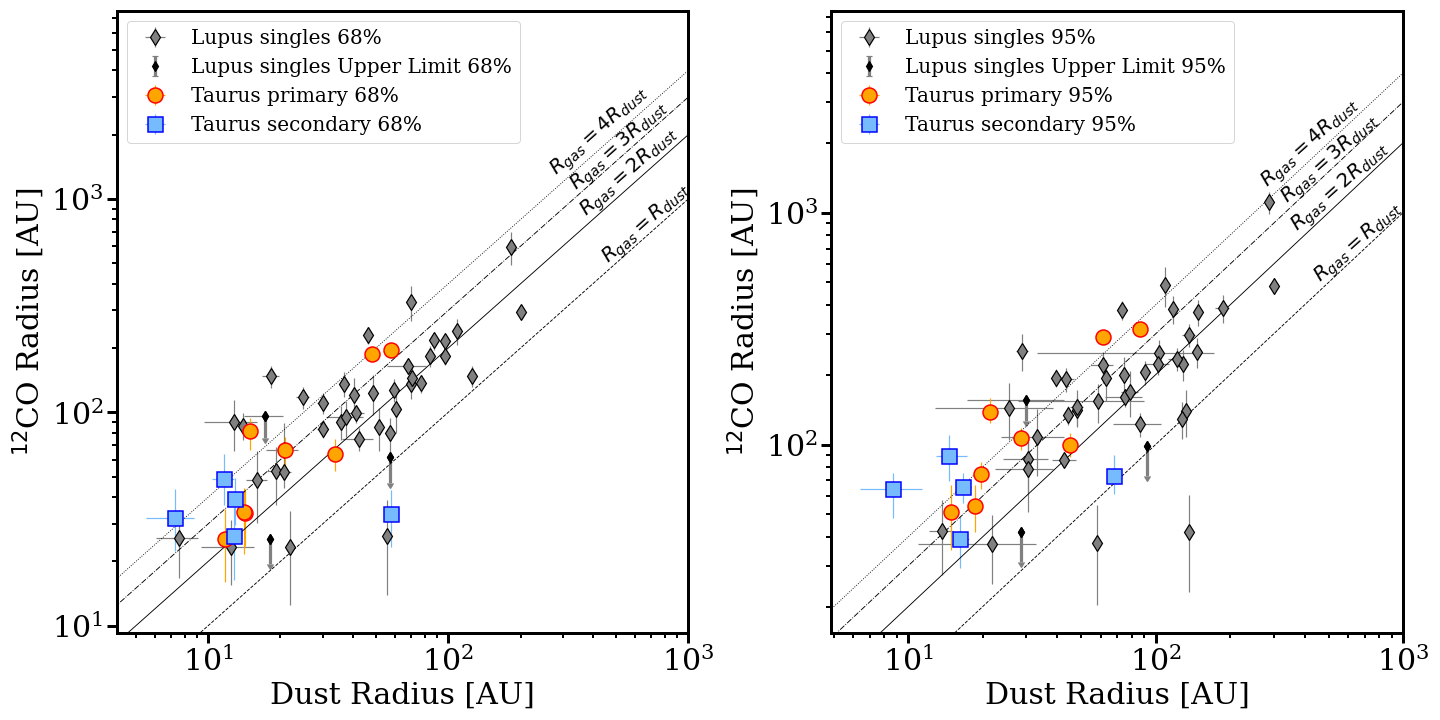}
	\caption{Comparison between the gas-to-dust size ratio for discs in the Taurus binaries (red and blue triangles) and for single discs in Lupus (grey diamonds). Effective disc radii $R_{68}$ (left) and disc radii $R_{95}$ (right) are shown. }\label{fig:LupusVSTaurusRatio}
\end{figure*}

Figures \ref{fig:LupusVSTaurusRatio} and \ref{fig:cdfratio} show the comparison between the 95\% and the 68 \% gas-to-dust disc size ratios measured in our sample of multiple stellar systems in the Taurus region, and the one measured by \citet{2021Sanchis+} for 42 discs in the Lupus star-forming regions, which are mainly singles or, in two cases, in wide binary systems with separation $\sim1000$ au, high enough that discs are not affected by tidal truncation \citep{2020Pearce+}.

The effective ratio $R_{68, \mathrm{gas}}/R_{68, \mathrm{dust}}$ estimated in Taurus multiple system is in very good agreement with the average ratio obtained in the sample of discs around single stars by \citet{2021Sanchis+} ($\sim 3.2$ vs $\sim 2.8$, respectively). The K-S test performed on these two samples confirms that the gas-to-dust size ratios in Taurus multiples are likely to be drawn by the same distribution of Lupus single discs (p-value $\sim 0.13$). On the contrary, the $R_{95, \mathrm{gas}}/R_{95, \mathrm{dust}}$ average ratio in Taurus multiple systems is larger than the ratio estimated by \citet{2021Sanchis+}, with the K-S test confirming the two samples to be statistically different (p-value $\sim 0.012$).

The difference in the values of $R_{95, \mathrm{gas}}/R_{95, \mathrm{dust}}$ estimated in discs around binaries and in singles is possibly due to the sharp truncation of the outer dusty discs in binary systems \citep{2019Manara+}. However, this difference may also be affected by the method through which the gas disc radii have been estimated. As discussed, we applied the cumulative flux technique directly on the zeroth moment image, with no parametric form to model the emission profile, while \citet{2021Sanchis+} modelled the CO emission of each disc by fitting the integrated line map to a Gaussian (or a Nuker) profile in the image plane, and then they applied the cumulative flux technique on the modelled emission profile. The latter method may lead to an underestimate of the 95\% disc radii, smoothing the outer emission of the discs.

In general, we conclude that the ratio of the gas to dust disc size in multiple stellar systems is found to be on the higher side of the distribution of values when compared to a population of more isolated systems and considering the 95\% disc radius, which is more sensitive to the full disc size.

\subsection{Comparison of gas disc radii with tidal truncation models}\label{sec:ecc}

We now compare the observed $^{12}$CO disc radii with the theoretical expectations for tidal truncation predicted by \cite{1994Artymowicz_Lubow}. As discussed by \citet{2019Manara+}, and references therein, it can be analytically computed that the truncation radius for a disc in a binary system with a circular orbit inclined along the plane of the sky (eccentricity $e=0$, mass ratio $q$ = 1, disc orbit misalignment $=0\degree$) is $\sim 0.33 ~a$, where $a$ is the semi-major axis of the binary orbit. The value of the truncation radius becomes larger for the circumprimary disc when the mass ratio $q$ is smaller than 1, and smaller for the circumsecondary disc for smaller values of $q$. The truncation radii instead decrease with higher eccentricities both in circumprimary and circumsecondary discs.
For systems with $e \gtrsim 0.9$, more violent processes than the disc-satellite interaction, such as the collision between the disc and the companion star, would occur.

 \begin{figure}[h]
    	\centering
	    \includegraphics[width=0.5\textwidth]{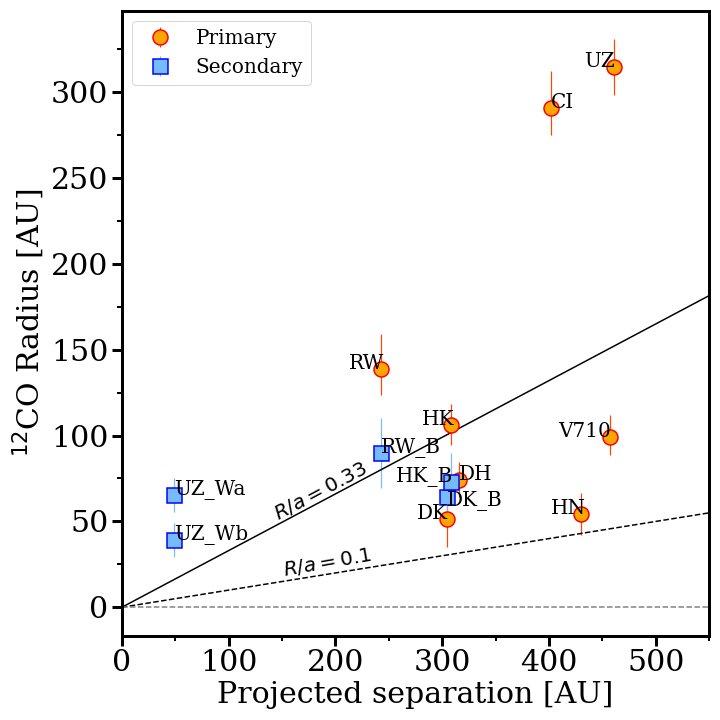}
	\caption{Observed $^{12}$CO gas radii as a function of the projected separation. Blue squares refer to the circumsecondary discs, while red circles to the circumprimary discs. }\label{fig:sep}
\end{figure}

Let us first assume that the projected separation $a_\mathrm{p}$ is $\sim a$, which again is correct if the orbit has low eccentricity and the plane of the orbit is aligned close to the plane of the sky, and compare the estimated gas disc radii $R_{95,gas}$ with the projected separation  between the components (reported in Table \ref{tab:ch4paramDisc}).
Since the UZ Tau system is a quadruple system (with UZ Tau E being a spectroscopic binary), when comparing observations with analytical models, we consider the UZ Tau system as two binaries. The first binary is composed by UZ Tau E and UZ Tau Wab (considering the sum of the masses of UZ Tau Wa and Wb), and has a projected separation of $3\farcs52$ (assuming the position of UZ Tau Wab as the center of mass of the two stars);  the second binary is composed by UZ Tau Wa and UZ Tau Wb, and has $a_\mathrm{p} =0\farcs375$. 
 
Figure \ref{fig:sep} shows that the typical observed ratio $R_{95,\mathrm{gas}}/a_\mathrm{p}$ in our sample is $\sim 0.15-0.35$, and more specifically $>0.1$ in all targets, contrary to the typical ratio of $\lesssim 0.1$ measured for the dust radii in \cite{2019Manara+}. This typical observed ratio is due to the fact that the gas disc radii are larger than the dust disc radii, as discussed in Section \ref{sec:dust-gas}, and already suggested in \cite{2019Manara+}. However, this is the first time that this value $R_{95,\mathrm{gas}}/a_\mathrm{p}$ is measured in a large sample of objects.

The observed distribution of gas disc radii points to the fact that the data do not agree with a simple description, as the ratio $R_{95,\mathrm{gas}}/a_\mathrm{p}$ typically differs from the expected value of 0.33 (only 2/13 discs show a ratio $\sim 0.33$). Therefore, a more detailed comparison is to be considered.

\begin{figure}[]
    	\centering
	    \includegraphics[width=0.5\textwidth]{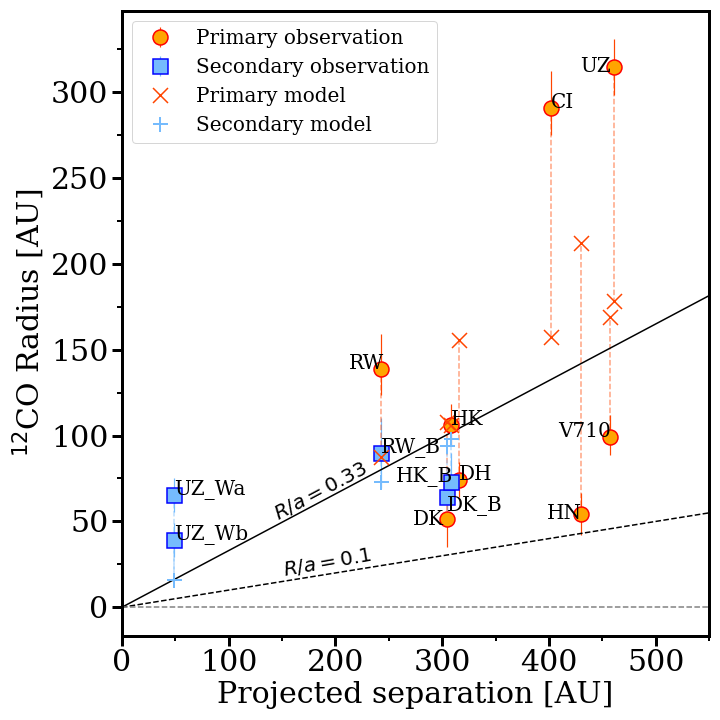}
	\caption{$^{12}$CO gas radii as a function of the projected separation. Blue squares and red circles show the observed ratio in the circumsecondary discs and the circumprimary discs, respectively. Blue pluses and red crosses show the expected ratio from tidal truncation model in the case of circular orbits ($e=0$) for circumsecondary discs and circumprimary discs, respectively. The longer the dashed line that links observed radii with models is, the less is the agreement with zero-eccentricity model. }\label{fig:sep0}
\end{figure}

To better quantify this result, we compare $R_{95,\mathrm{gas}}/a_\mathrm{p}$ with theoretical models by \cite{1994Artymowicz_Lubow}, using the equation derived by \citet{2019Manara+} under the assumption that the discs and the binary orbit are co-planar:
\begin{multline}\label{eq::trunc}
\frac{R_{\rm trunc}}{a_{\rm p}} = \frac{0.49 \cdot q^{2/3}_i}{0.6\cdot q_i^{2/3} + \ln(1+q_i^{1/3})}\left(b\cdot e^c + 0.88\mu^{0.01}\right) \cdot \\
\cdot \left[ \frac{1-e^2}{1+e\cdot \cos \nu}\sqrt{1-\sin^2(\omega+\nu)\sin^2i} \right]^{-1},
\end{multline}
where 
$\nu$  the true anomaly, $\omega$  the longitude of periastron, and \textit{i}  the inclination of the plane of the orbit with respect to the line of sight, $q_i$ is the mass ratio (either $q_1 = M_1/M_2$ or $q_2 = q = M_2/M_1$), and $b$ and $c$ are the parameters derived by \citet{2019Manara+} that depend on the disc viscosity or equivalently on the Reynolds number, $\rey$. 
Figure \ref{fig:sep0} shows this comparison in the case of zero-eccentricity orbits ($e=0$) and inclination of the orbit $i=0$. Red circles and blue squares show the observed ratio for circumprimary and circumsecondary discs, respectively. Red crosses and blue pluses show for each target the ratio expected from model in the case $e=0$. The longer the dashed line that links observed radii with models is, the less is the measured ratio in agreement with zero-eccentricity model predictions. Excluding the discs around UZ Tau E and CIDA 9 A and the disc around RW Aur A, the measured values of the observed ratios point to disc radii that are typically smaller than expectations from tidal truncation models at zero eccentricity.

\begin{figure*}[h]
    	\centering
	    \includegraphics[trim=0.cm 2cm 0cm 0.55cm,clip,width=1.\textwidth]{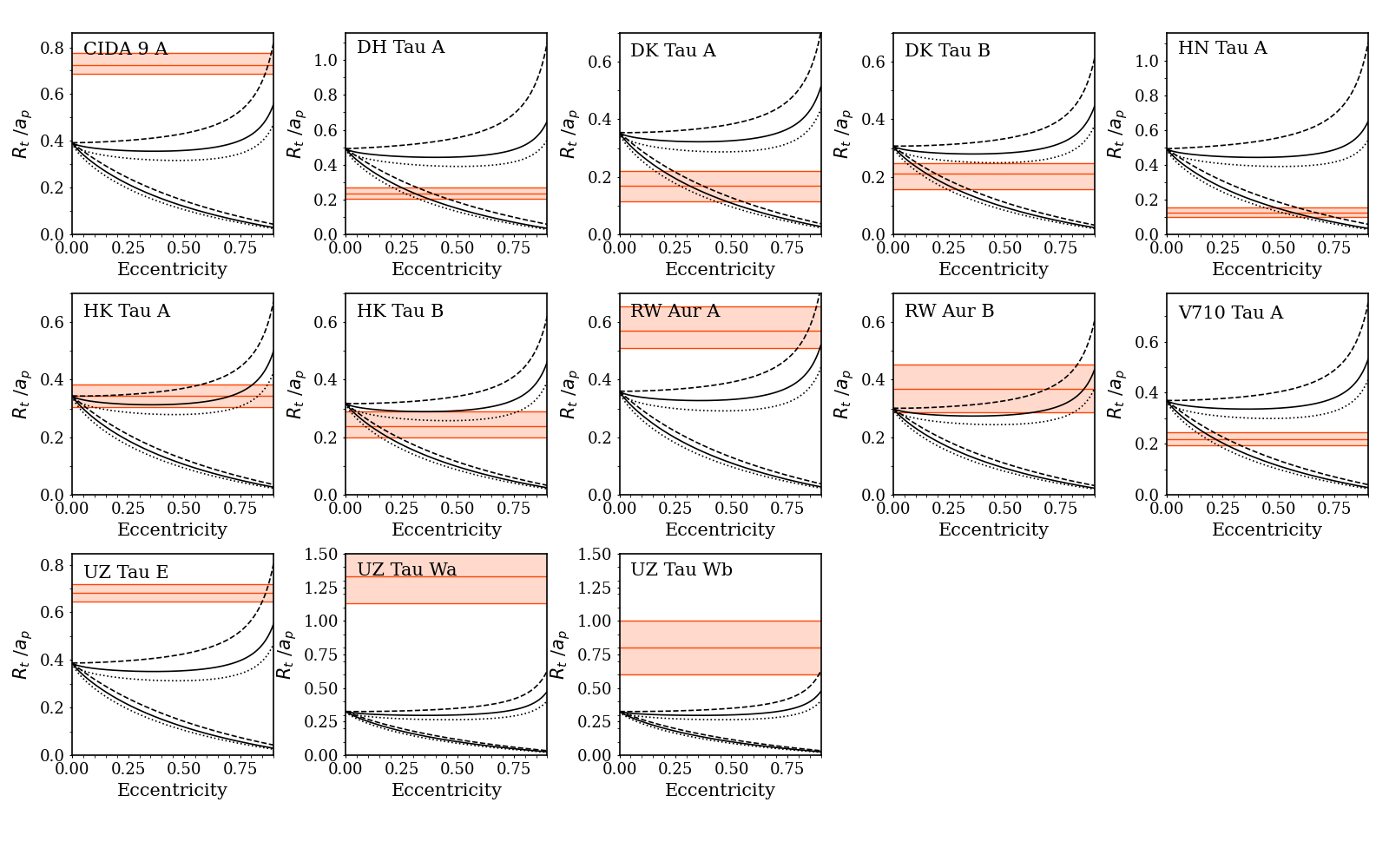}
	\caption{Condition on the eccentricity of the stellar orbits for which the estimated truncation radii of the discs in the sample are in agreement with theoretical predictions (see Sect.~\ref{sec:ecc}). Red bands show the estimated $R_{95, \mathrm{gas}}/a_\mathrm{p}$ with their uncertainties. Black curves show the expected truncation radius in unit of the projected separation for different Reynolds numbers -- dashed curves for $\mathcal{R} = 10^4$, solid for $\mathcal{R}= 10^5$, and dotted for $\mathcal{R}= 10^6$. The lower black curves are obtained assuming that the secondary is observed at apoastron, while black curves at the top assuming that the secondary is located at periastron. The expected $R_\mathrm{t}/a_\mathrm{p}$ increases for higher orbital inclination, and thus the black curves would move to higher values in the plot if the assumption of face-on orbits is not fulfilled (see Equation \ref{eq:eq2}).}\label{fig:eccMod}
\end{figure*}

As discussed in \cite{2019Manara+}, at a given inclination of the orbital plane $i$, the truncation radius $R_\mathrm{t}$ in units of the projected separation $a_\mathrm{p}$ has a minimum when the secondary star is observed at the apoastron $-$ $a_\mathrm{p} = a(1+e)$ $-$ and has a maximum at the periastron $-$ $a_\mathrm{p} = a(1-e)$:
\begin{equation}\label{eq:eq2}
    \frac{R_\mathrm{t}}{a}\frac{1}{(1+e)\cos i} < \frac{R_\mathrm{t}}{a_\mathrm{p}} < \frac{R_\mathrm{t}}{a}\frac{1}{(1-e)\cos i}.
\end{equation}
The minimum ratio is found at $i=0 \si{\degree}$ and $R_\mathrm{t}/a_\mathrm{p}$ increases for higher orbital inclination. In Figure~\ref{fig:eccMod}, we plot the ratio of the truncation radius to the projected separation of the orbit as a function of eccentricity for all the targets in the sample, assuming an orbital inclination $i=0 \si{\degree}$ with respect to the line of sight. 
The intersection between the observed $R_\mathrm{t}/a_\mathrm{p}$ (red bands in Figure~\ref{fig:eccMod}) and the theoretical model assuming that the secondary is observed at apoastron (lower black curves in the figure) represents the minimum possible value of the eccentricity for the system. This statement holds  even in case that the orbital plane of the binary is misaligned with respect to the disc ($i>0 \si{\degree}$), since the black curves in Figure~\ref{fig:eccMod} will move to higher values leading to a larger minimum eccentricity.
As an example, we refer to the V710 A disc (second row, last column in the figure). To be compatible with tidal truncation models, the observed ratio of disc radius to separation must lie between the bottom and the top black curves. The observed ratio is shown as a red band in the plot, from which we see that, in this case, the eccentricity must be larger than $\simeq 0.2$. For inclined orbits, the minimum eccentricity must be even larger.
The dominant uncertainty on the eccentricity estimated in this way comes from the lack of information on the local Reynolds number, and hence different Reynolds numbers are assumed -- $\mathcal{R} = 10^4, 10^5, 10^6$.

\begin{figure}[h]
    	\centering	    \includegraphics[width=0.5\textwidth]{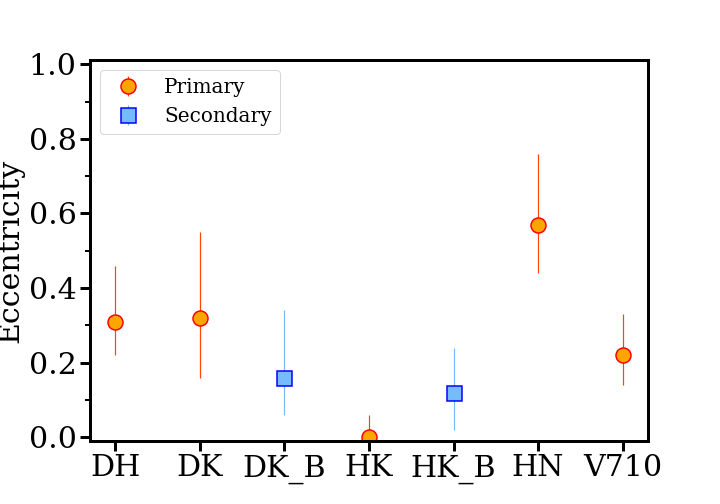}
	\caption{Estimated minimum eccentricities for each target. Red circles refer to circumprimary discs, while blue squares refer to circumsecondary discs. }\label{fig:eccValues}
\end{figure}

For each target, Table \ref{tab:ecc} shows the ratio between disc radii and separations predicted by the models assuming $e=0$ and
the observed ratio between disc radii $R_{\mathrm{95,gas}}$ and projected separations. Finally, the minimum eccentricities of the orbits estimated assuming zero orbital inclination and physical separation $a$ between the components that matches the projected separation $a_\mathrm{p}$ are shown in the last column in Table \ref{tab:ecc} and in Figure \ref{fig:eccValues}.

\begin{table}[h]
    \centering
   
     \caption{Comparison between observed disc radii and tidal truncation models.}
    
    \begin{tabular}{lccc}
       
        Target &  $(R_\mathrm{disc}/a )\mathrm{by L\&A94}$ & $R_\mathrm{95,gas}/a_{ \mathrm{p}}$ & $e_\mathrm{min}$ \\
        \hline\hline

CIDA 9 A & $0.39$ & $0.72$ & - \\
DH Tau A  & $0.49$ & $0.24$ & $0.31^{+0.15}_{-0.09}$ \\ 
DK Tau A & $0.35$ & $0.17$ & $0.32^{+0.23}_{-0.16}$  \\
DK Tau B & $0.31$ & $0.21$ & $0.16^{+0.18}_{-0.10}$ \\
HK Tau A & $0.34$ & $0.34$ & $0.00^{+0.06}_{-0.00}$ \\
HK Tau B & $0.32$ & $0.24$ & $0.12^{+0.12}_{-0.10}$  \\
HN Tau A & $0.49$ & $0.13$ & $0.57^{+0.19}_{-0.13}$ \\
RW Aur A & $0.36$ & $0.57$ & 0$^\ddag$ \\
RW Aur B & $0.30$ & $0.37$ & 0$^{\dagger}$\\
UZ Tau E & $0.39$ & $0.68$ & - \\
UZ Tau Wa & $0.34$ & $1.33$ & - \\
UZ Tau Wb & $0.33$ & $0.80$ & 0$^*$ \\
V710 Tau A & $0.37$ &  $0.22$ & $0.22^{+0.11}_{-0.08}$ \\

\hline
 \end{tabular}
  \tablefoot{Second column: expected ratio between disc radii and separation as predicted by \cite{1994Artymowicz_Lubow} when assuming circular orbits ($e=0$). Third column: observed ratio between disc radii $R_{\mathrm{95,gas}}$ and projected separation. Fourth column: minimum eccentricities of the orbits estimated assuming zero orbital inclination and physical separation $a$ between the components that matches the projected separation $a_\mathrm{p}$.
  Agreement with analytical model with zero eccentricity  $^{\dagger}$ within one sigma; $^*$ within 1.5 sigmas; $^\ddag$ within 3.5 sigmas (see Figure~\ref{fig:eccMod}).}
    \label{tab:ecc}
\end{table}

The tidal truncation models assuming face-on inclination of the orbit cannot explain the observed large values of $R_\mathrm{t}/a_\mathrm{p}$ in the CIDA 9, UZ Tau, and RW Aur systems, except assuming unrealistic parameters ($e\sim 1$ and targets observed at the periastron); this can be explained either by a high inclination of the orbit with respect to the plane of the sky and by the fact that the disc sizes are regulated by other additional processes (e.g. substructures and interaction between the components) -- not included in the model -- on top of tidal truncation.
Excluding these systems, the estimated minimum eccentricities show typical values of $\sim 0.15-0.5$ and an average value of $\sim 0.3$, in very good agreement with the observed distribution of eccentricities for main-sequence binary systems of low-mass \citep[e.g.,][]{2013Duchene_Kraus}. This confirms observationally the result by \citet{zagaria21b} that no highly eccentric orbits must be invoked to explain the observed small dust radii of discs in multiple stellar systems.
When comparing both the circumprimary and circumsecondary disc sizes to the models -- comparison possible only in DK Tau and HK Tau systems, a very good agreement with the two estimated eccentricities is found.

\begin{figure}[h]
    	\centering	    \includegraphics[width=0.5\textwidth]{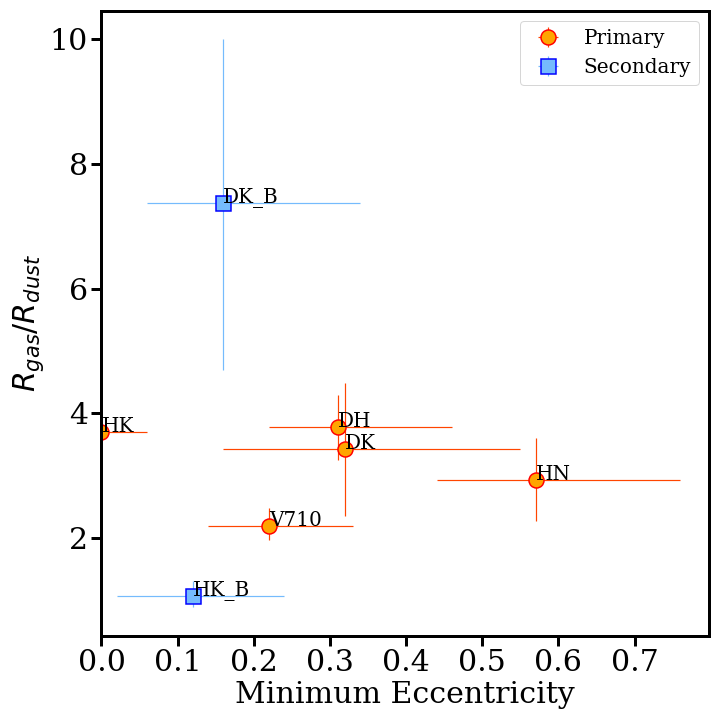}
	\caption{Gas-dust size ratios ($R_\mathrm{95,gas}/R_\mathrm{95,dust}$) as a function of the minimum eccentricities of the orbits. Red circles refer to circumprimary discs, while blue squares refer to circumsecondary discs.}\label{fig:ratioVSecc}
\end{figure}

We finally explore whether more eccentric orbits would have an impact on the relative sizes of the gas and dust components of discs, as predicted by previous theoretical works \citep[e.g.,][]{1993Clarke_Pringle}. Figure \ref{fig:ratioVSecc} shows that no correlation is observed between the observed gas-dust size ratios and the estimated eccentricity of the orbits. However, to definitively refute a dependence of the gas-dust size ratio on the eccentricity, it is necessary to significantly increase the sample of systems with measured size ratio and eccentricity.

\subsection{Disc misalignment}

\begin{figure}[h]
    	\centering	    
    	\includegraphics[width=0.5\textwidth]{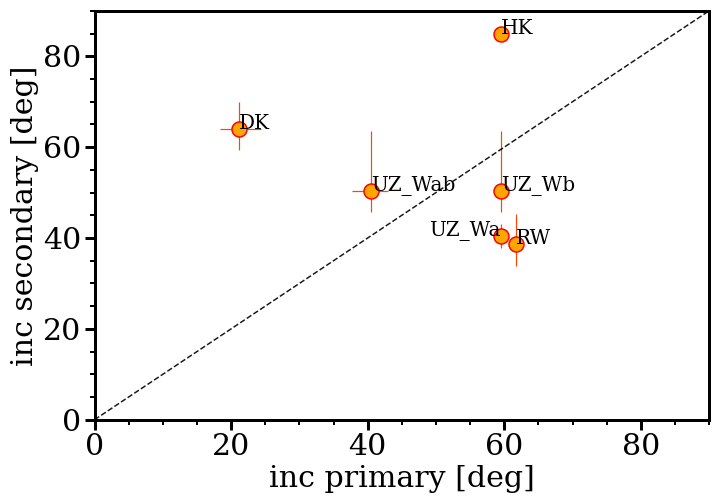}
    	\\
    	\includegraphics[width=0.5\textwidth]{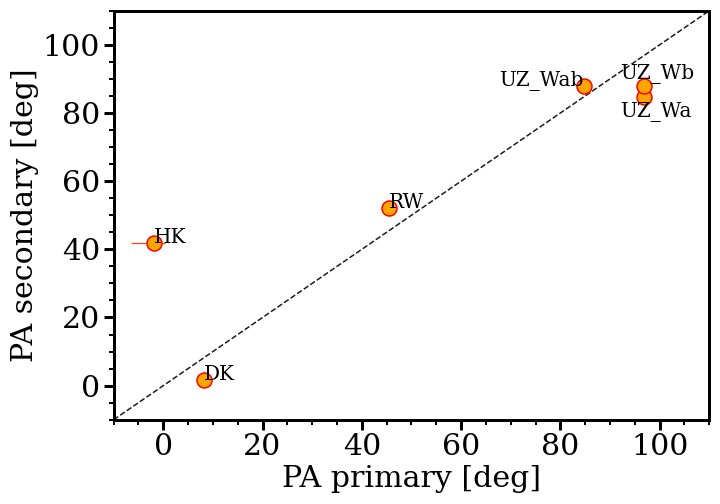}

	\caption{Comparison between the position angles (top) and the inclination (bottom) for the discs around the primary and the secondary in each multiple stellar system where both discs are detected and resolved. In the case of UZ Tau system, we considered UZ Tau system as a two binaries, comparing UZ Tau E parameters with UZ Tau Wab parameters (comparison labeled as UZ\_Wa and UZ\_Wb for the two UZ Tau Wab discs) and UZ Tau Wa parameters with UZ Tau Wb parameters (comparison labeled as UZ\_Wab). In the other cases, the first two letters of the target name are labelled. }\label{fig:misal}
\end{figure}

\begin{figure}[h]
    	\centering	    \includegraphics[width=0.4\textwidth]{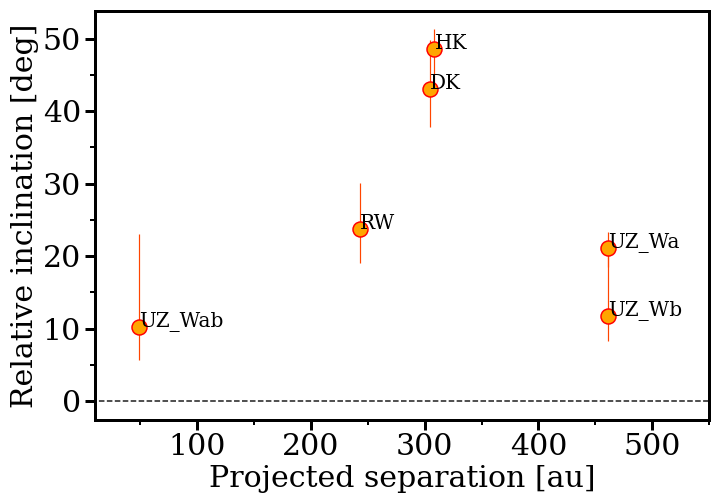}
    
	\caption{Relative inclination of the discs in each multiple stellar systems as a function of their projected separation. In the case of UZ Tau system, we considered UZ Tau system as a two binaries, comparing UZ Tau E parameters with UZ Tau Wab parameters (comparison labeled as UZ\_Wa and UZ\_Wb for the two UZ Tau Wab discs) and UZ Tau Wa parameters with UZ Tau Wb parameters (comparison labeled as UZ\_Wab). In the other cases, the first two letters of the target name are labelled. }\label{fig:relInc}
\end{figure}

The amount of alignment of the plane of rotation of disc in a multiple system can be used as a constraint to star formation models. In the most simplified case, disc fragmentation would tend to form close binaries with aligned angular momentum vectors, while turbulent fragmentation would produce more randomly distributed binary orientation \citep[e.g., ][]{2000Bate,2010Offner+, 2010Kratter+, 2018Bate}. It is therefore instructive to explore the
observed values of the relative inclinations for the discs in our sample to be used to constrain disc evolution models in multiple systems.

Figure~\ref{fig:misal} shows the alignment of discs around the components of the multiple systems in which both components are detected and resolved. The position angles and inclinations used to make this comparison are those reported in Table \ref{tab:estimatedDiscProp}. The first two letters of the system name are labeled. In the case of the UZ Tau system, we compare the UZ Tau E disc with the UZ Tau Wab discs, separately (labeled as UZ\_Wa and UZ\_Wb), and the UZ Tau Wa disc with the UZ Tau Wb disc.
The inclination of the components of the multiple systems appear to be different in most cases (top panel), with only the East and West components of UZ~Tau showing some agreement. 
The position angles of the discs are also usually similar, with the exception of the HK Tau discs, which show different values (bottom panel).

However, the degree of alignment should be explored using the relative inclination of the discs, computed using the spherical law of cosines: $\cos(\Delta i)=\cos(i_1)\cdot \cos(i_2)+\sin(i_1) \sin(i_2) \cos(\Omega_1 - \Omega_2)$, with $i$ the inclination (ambiguous in sign, so that it leads to two families of solutions) and $\Omega$ the position angle. The result is shown in Figure~\ref{fig:relInc}. The discs in the UZ Tau system show relative inclination differences $\lesssim 0-20 \si{\degree}$, while the RW Aur, DK Tau and HK Tau systems show a larger relative inclination ($\sim 25 \si{\degree}$, $\sim 40 \si{\degree}$, and $\sim 50 \si{\degree}$, respectively). For the latter, this is in line with the earlier results of \citet{2014Jensen_Akeson}. We conclude that most of the observed binaries appear not to be coplanar, possibly a result of turbulent fragmentation, rather than disc fragmentation.

\section{Conclusions}\label{sect:conclusions}

 We have presented here new Band 6 ALMA observations of CO line and continuum emission in ten multiple stellar systems in the Taurus star-forming region. The observed sample comprises seven binaries, one triple, one quadruple and one star part of a very wide binary system, and was observed at a spatial resolution of $\sim 0\farcs15$ and with an integration time of $\sim 40$ min/target. Three systems (GK Tau, UY Aur and T Tau) were not analysed here. One additional multiple system, RW~Aur, was also included in the analysis using ALMA archival data. Of this sample, eight multiple stellar systems were analysed in this work.
 
 We derived the disc geometrical properties using the \textit{eddy} tool \citep{2019Teague}, and then performed an image plane analysis of the data, to estimate the disc total fluxes with a cumulative flux technique. Assuming as disc radii those enclosing a certain fraction of the total disc fluxes (68\% or 95\%), we compared the gas radii to the dust radii estimated by \citet{2019Manara+}, and derived the gas-to-dust size ratio to be $\sim 2-4$. The effective (68\%) gas-to-dust size ratio distribution is found to be statistically compatible to the ratio estimated by \citet{2021Sanchis+} in a population of single discs. On the contrary, considering the 95\% disc radius, the gas-to-dust size ratio is found to be  on the high-end of the distribution of the gas-dust size ratios measured in a population of more isolated systems, possibly due to the sharp truncation of the outer dusty discs in binary systems \citep{2019Manara+}.
 
 We compared our estimates with analytical predictions for the tidal truncation on disc sizes in binary systems \citep{1994Artymowicz_Lubow}. In general, the 95\% gas disc radii are $\sim 0.15-0.35$ times the projected separation of the binaries, suggesting that the systems are not in circular orbits. 
 Exception to this typical ratio are the discs around UZ Tau E, CIDA 9 A, and RW Aur that show very large value of this ratio possibly due to projection effects or additional processes regulating disc truncation (e.g. substructure for the first two systems and interaction between the component for RW Aur).
 When comparing these ratios with the theoretical predictions, the minimum eccentricities of the orbits are $0 \lesssim e \lesssim 0.5$, with an average value of $\sim 0.3$, in good agreement with expectations \citep[e.g.,][]{2013Duchene_Kraus}. Finally, the discs in multiple systems appear to be misaligned, possibly a result of turbulent fragmentation.
 
 This study shows the importance of deep ALMA observations of line emission from discs in multiple stellar systems to derive their sizes and constrain models of tidal interactions. 
 Future studies targeting multiple systems should aim at larger samples and should cover a larger span of binary separations, in particular focusing on the shorter separations. 
 
\begin{acknowledgements}
We thank the anonymous referee for the comments that greatly improved this paper.
This paper makes use of the following ALMA data: ADS/JAO.ALMA\#2018.1.00771.S, ADS/JAO.ALMA\#2016.1.00877.S, ADS/JAO.ALMA\#015.1.01506.S. ALMA is a partnership of ESO (representing its member states), NSF (USA) and NINS (Japan), together with NRC (Canada), MOST and ASIAA (Taiwan), and KASI (Republic of Korea), in cooperation with the Republic of Chile. The Joint ALMA Observatory is operated by ESO, AUI/NRAO and NAOJ. The authors are thankful to Joey Rodriguez for sharing the reduced CO cubes for RW Aur, to Enrique Sanchis for sharing the values of the gas disc radii prior to publication, and to Leonardo Testi for very helpful tips on the self-calibration of the data. A.A.R. thanks Cristiano Longarini for useful discussions.
A.A.R. acknowledges support from ESO for a six month visit through the SSDF Funding. This project has received funding from the European Union's Horizon 2020 research and innovation programme under the Marie Sklodowska-Curie grant agreement No 823823 (DUSTBUSTERS).
This work was partly supported by the Deutsche Forschungs-Gemeinschaft (DFG, German Research Foundation) - Ref no. FOR 2634/1 TE 1024/1-1.
M.T. has been supported by the UK Science and Technology research Council (STFC) via the consolidated grant ST/S000623/1, and by the European Union’s Horizon 2020 research and innovation programme under the Marie Sklodowska-Curie grant agreement No. 823823 (RISE DUSTBUSTERS project). FMe, GvdP acknowledge funding from ANR of France under contract number ANR-16-CE31-0013.
H.-W.Y. acknowledges support from Ministry of Science and Technology (MOST) in Taiwan through the grant MOST 108- 2112-M-001-003-MY2.
P.P. acknowledges support provided by the Alexander von Humboldt Foundation in the framework of the Sofja Kovalevskaja Award endowed by the Federal Ministry of Education and Research.
E.R. acknowledges financial support from the European Research Council (ERC) under the European Union's Horizon 2020 research and innovation programme (grant agreement No. 681601 and No. 864965).
F.L. acknowledges support from the Smithsonian Institution as the Submillimeter Array (SMA) Fellow. D.H. is supported by CICA through a grant and grant number 110J0353I9 from the Ministry of Education of Taiwan.

\end{acknowledgements}

\bibliography{bibliography.bib}

\appendix
\section{Maps and spectra for all sample}\label{sec:spectra+maps}

In this section, we report the maps, the spectra, and the channel maps of all analysed and detected targets. Although in this work we focus on the $^{12}$CO emission, the targeted emission lines include also CO isotopologues $-$ $^{13}$CO and C$^{18}$O.

\begin{figure*}[]
    	\centering
	    \includegraphics[width=\textwidth]{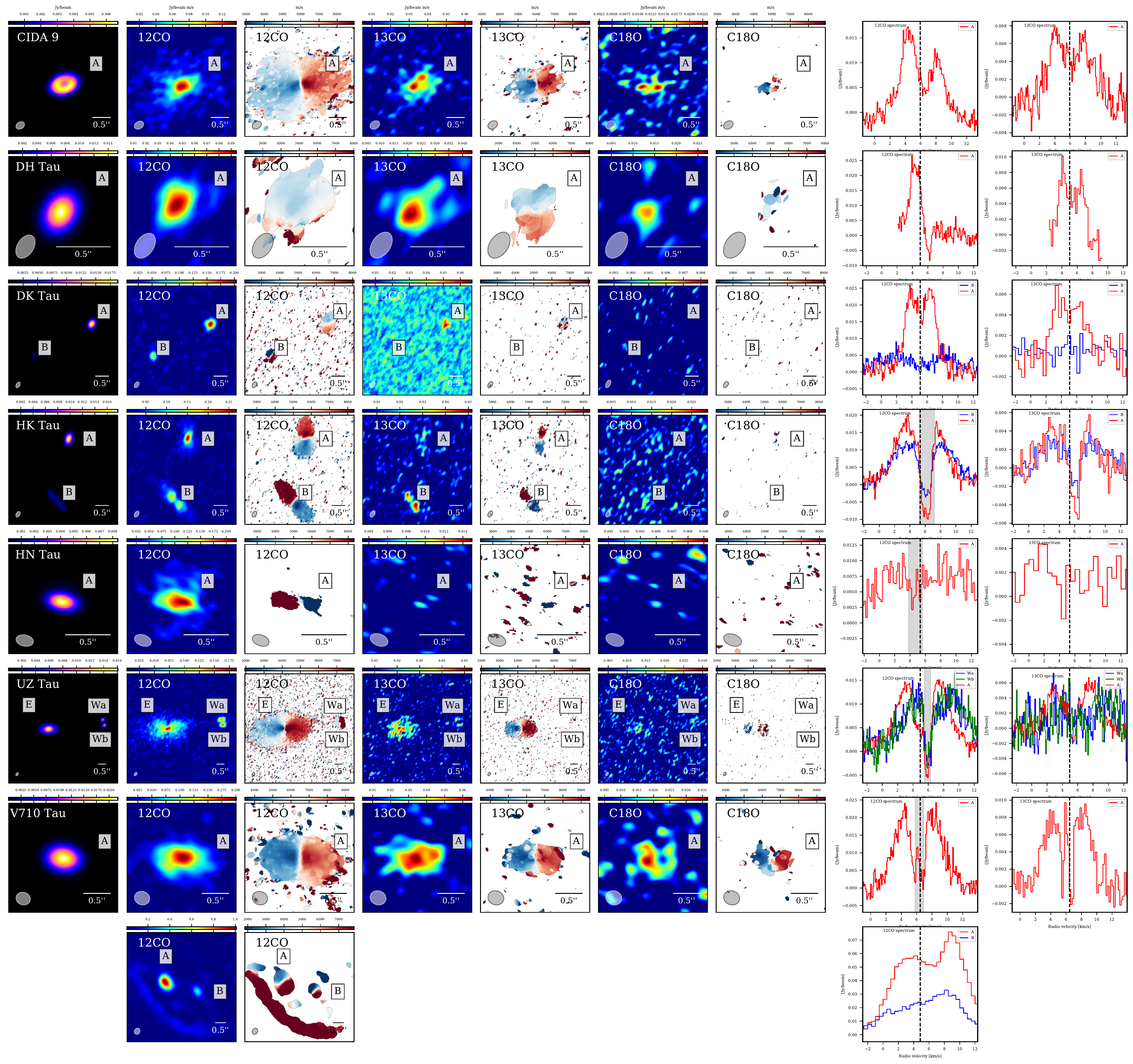}
	\caption{Images and spectra of all target in the analysed sample. Each row refers to a system in the sample. The first columns show the continuum images with the names of the targets indicated in the top left. The second, fourth and sixth columns show the zeroth moment images for the $^{12}$CO, the $^{13}$CO, and the C$^{18}$O emission, respectively  -- scaled so that the maximum is equal to the peak flux and the minimum is clipped at the image rms. The third, fifth and seventh columns show the first moment images (calculated through the \textit{bettermoments} tool) for the $^{12}$CO, the $^{13}$CO, and the C$^{18}$O emission, respectively -- scaled so that the maximum and the minimum are equal to the systemic velocity of the target $\pm 2$ km/s. The FWHM beam size is shown in the bottom left of each panel. All bars in the bottom right of each panel are 0\farcs5 long. The last two columns show the $^{12}$CO and $^{13}$CO spectra of the circumprimary (red), circumsecondary (blue), and circumtertiary (green) discs (when applicable), with vertical black lines showing the systemic velocities fitted with the \textit{eddy} tool and grey band showing the excluded absorbed velocities in \textit{eddy} fitting on circumprimary disc rotation maps (see Table~\ref{tab:cloudAbs}).}\label{fig:maps}
\end{figure*}

The continuum and CO zeroth and first moment maps are shown with the integrated spectra for all targets in Figure~\ref{fig:maps}. The detected discs (see Table~\ref{tab:detections}) show a rotation pattern, that is fitted with \textit{eddy} (Section~\ref{sec:eddy}). 

\begin{figure*}[]
    	\centering
	    \includegraphics[width=0.98\textwidth]{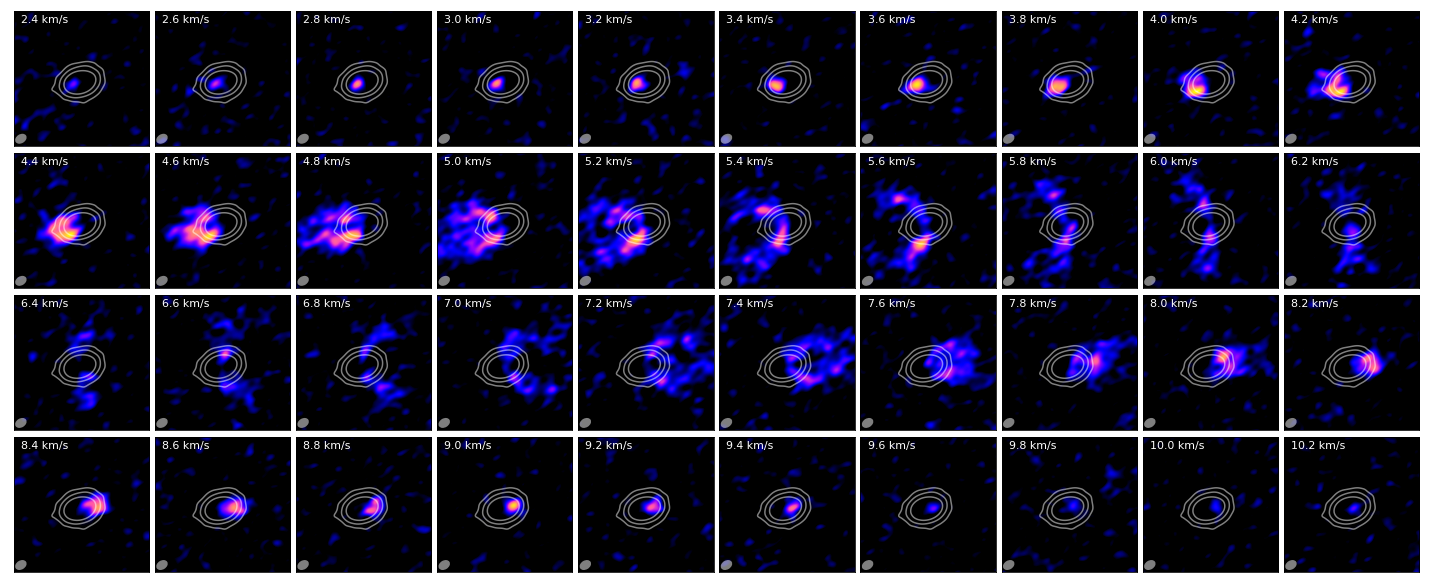}
	\caption{Channel maps of the $^{12}$CO emission in CIDA 9 A disc. Each panel shows the velocity of the channel in the upper left corner. The FWHM beam size is shown in the bottom left. The images are scale so that the maximum is equal to the peak flux and the minimum is clipped at the image rms. White contours show 5, 30, and 100 times the rms of the continuum emission. }\label{fig:CIDA9_chmaps}
\end{figure*}

\begin{figure*}[]
    	\centering
	    \includegraphics[width=0.98\textwidth]{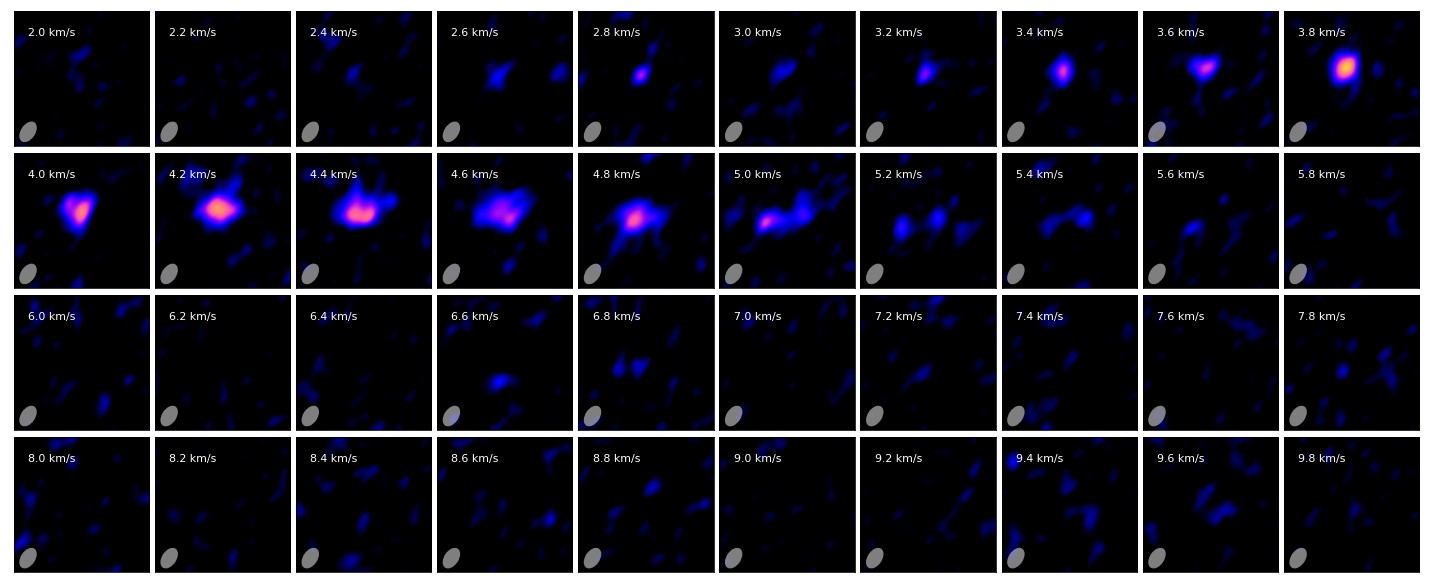}
	\caption{Channel maps of the $^{12}$CO emission in DH Tau A disc. Same panels and symbols as in Figure~\ref{fig:CIDA9_chmaps}. }\label{fig:DHTau_chmaps}
\end{figure*}

\begin{figure*}[]
    	\centering
	    \includegraphics[width=0.98\textwidth]{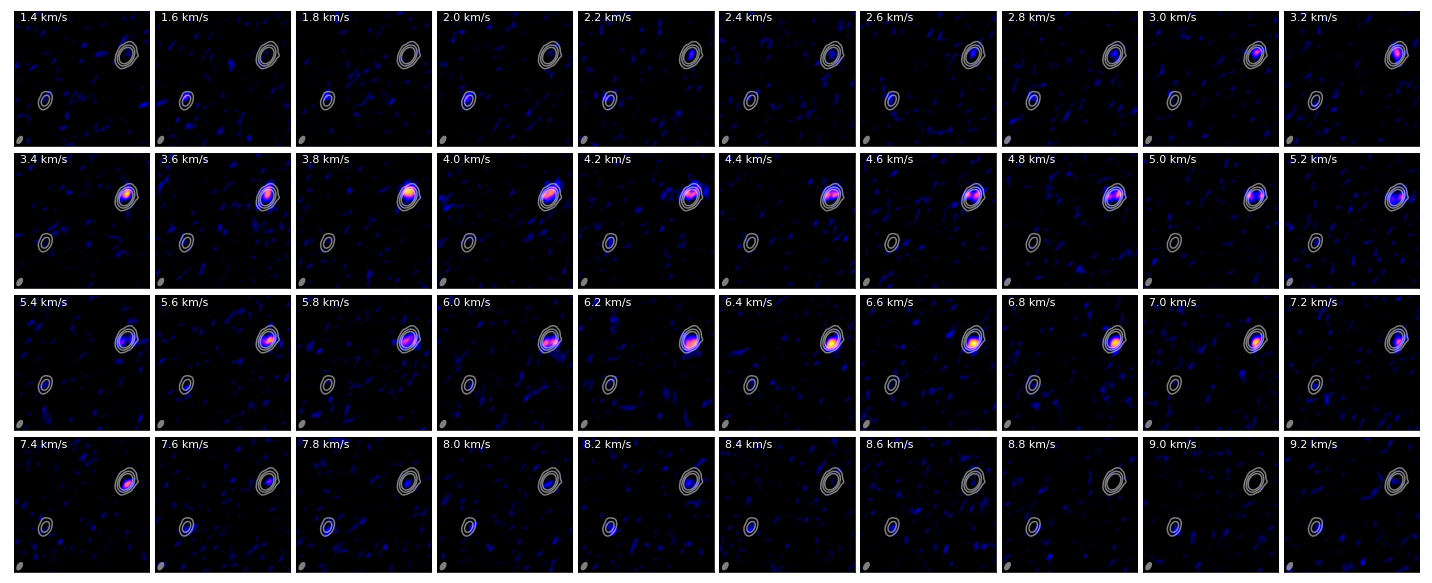}
	\caption{Channel maps of the $^{12}$CO emission in DK Tau discs. Same panels and symbols as in Figure~\ref{fig:CIDA9_chmaps}. }\label{fig:DKTau_chmaps}
\end{figure*}

\begin{figure*}[]
    	\centering
	    \includegraphics[width=0.98\textwidth]{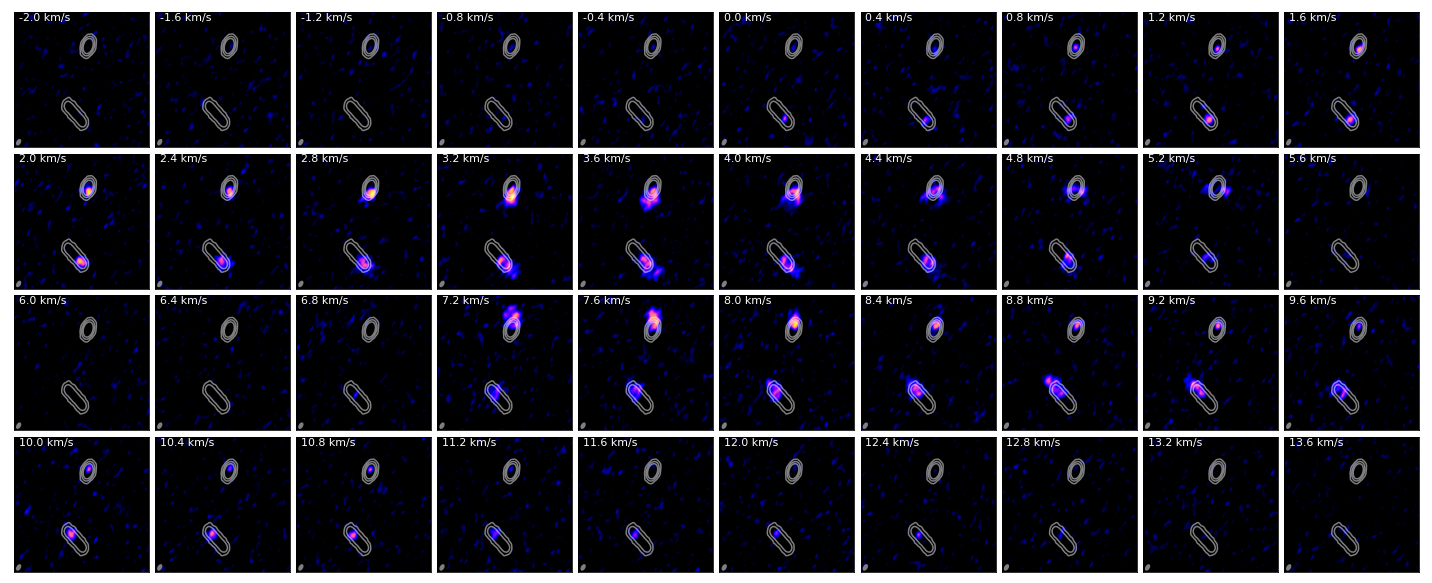}
	\caption{Channel maps of the $^{12}$CO emission in HK Tau discs. Same panels and symbols as in Figure~\ref{fig:CIDA9_chmaps}. }\label{fig:HKTau_chmaps}
\end{figure*}

\begin{figure*}[]
    	\centering
	    \includegraphics[width=0.98\textwidth]{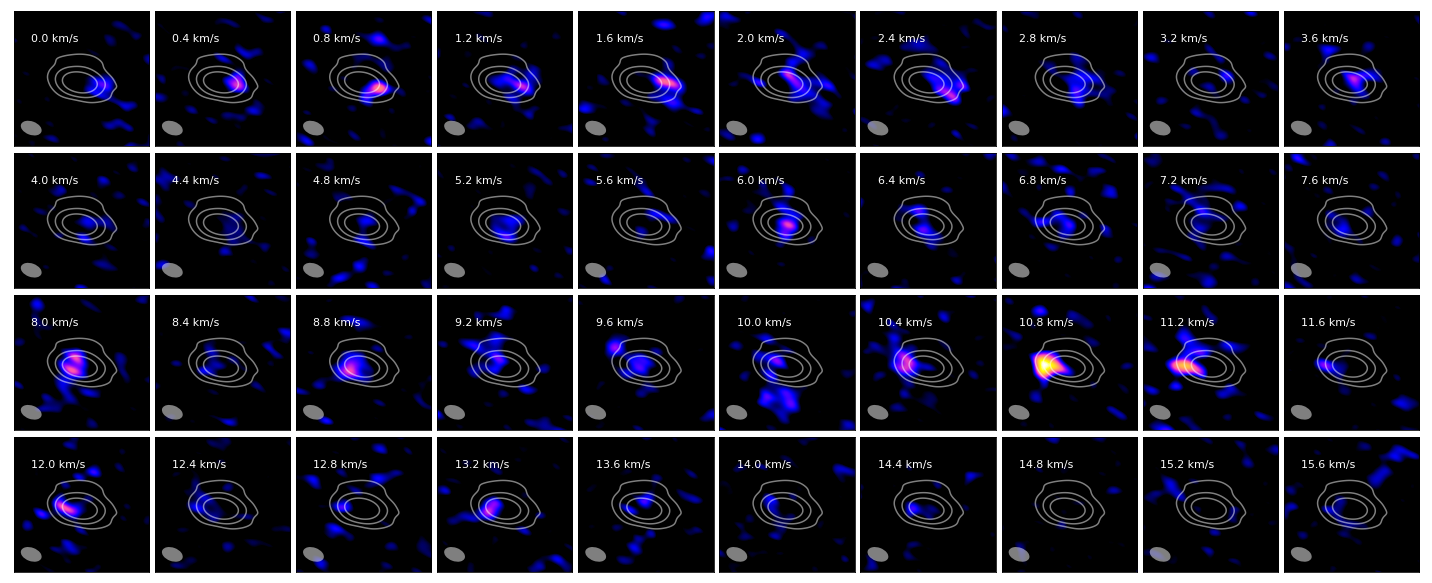}
	\caption{Channel maps of the $^{12}$CO emission in HN Tau A disc. Same panels and symbols as in Figure~\ref{fig:CIDA9_chmaps}. }\label{fig:HNTau_chmaps}
\end{figure*}

\begin{figure*}[]
    	\centering
	    \includegraphics[width=0.98\textwidth]{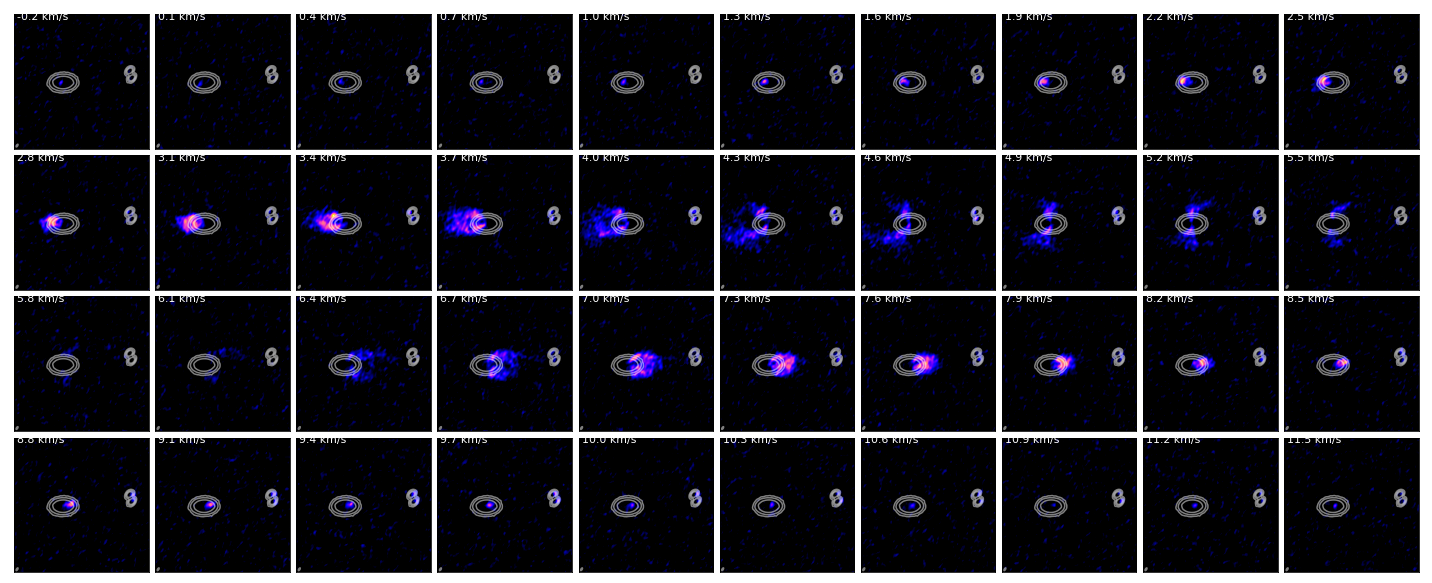}
	\caption{Channel maps of the $^{12}$CO emission in UZ Tau discs. Same panels and symbols as in Figure~\ref{fig:CIDA9_chmaps}. }\label{fig:UZTau_chmaps}
\end{figure*}

\begin{figure*}[]
    	\centering
	    \includegraphics[width=0.98\textwidth]{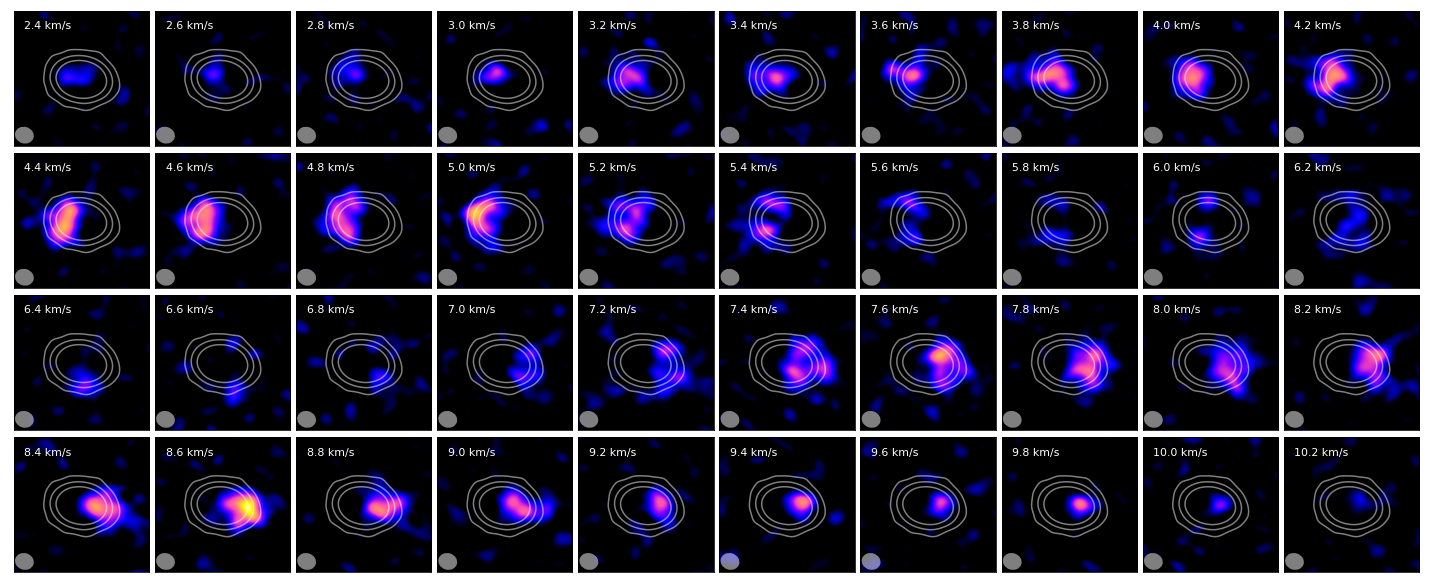}
	\caption{Channel maps of the $^{12}$CO emission in V710 Tau A disc. Same panels and symbols as in Figure~\ref{fig:CIDA9_chmaps}. }\label{fig:V710Tau_chmaps}
\end{figure*}

The channel maps of shown in Figures~\ref{fig:CIDA9_chmaps}-\ref{fig:V710Tau_chmaps} are used also to identify the channels affected by absorption due to the foreground molecular cloud, particularly pronounced in the $^{12}$CO emission. The velocity ranges affected by cloud absorption are reported in Table~\ref{tab:cloudAbs}, and excluded from the analysis with \textit{eddy}.


  \section{Comparison between new continuum data and older continuum data}\label{app:comparison_new_old}
  
As described in Sect.~\ref{sect:sample}, the new data analyzed here have been taken for the same sample as the one presented by \citet{2019Manara+} and \citet{2019Long+}. The main differences of these samples, on top of the presence in the new one of the $^{12}$CO emission, are the spatial resolution and the sensitivity of the observations. The new data have slightly worst spatial resolution ($0\farcs15$ vs $0\farcs12$) and better sensitivity ($\sim$30 $\mu$Jy rms vs $\sim50 \si{\micro}$Jy beam$^{-1}$). Whereas this work focuses on the analysis of the gas emission, we discuss here the differences between the two data-sets in the continuum emission, and motivate the choices of which data-sets and analysis techniques have been used to discuss our results.

\begin{figure*}[h]
    	\centering
	    \includegraphics[width=\textwidth]{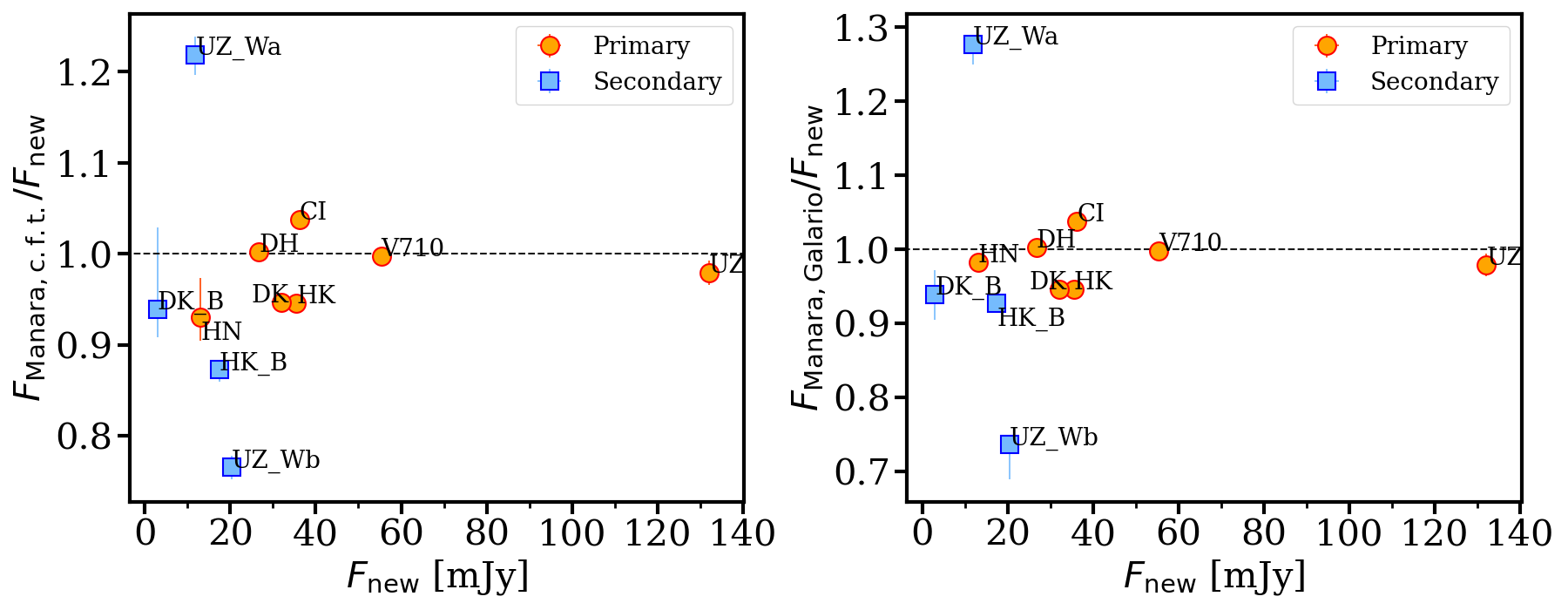}
	\caption{Comparisons between the measured total disc fluxes on the new continuum observation and on the data presented by \cite{2019Manara+}. The left panel shows the ratio between the total fluxes calculated on the data by \cite{2019Manara+} ($F_\mathrm{Manara,c.f.t.}$) and on the new continuum data ($F_\mathrm{new}$), both estimated through the cumulative flux technique. The right-hand panel, instead, reports the ratio using the fluxes calculated by \cite{2019Manara+} with the uv-modelling method ($F_\mathrm{Manara,Galario}$), and on the new continuum data with the cumulative flux technique. Red circles are used for circumprimary discs, blue squares for circumsecondary ones.}\label{fig:flux}
\end{figure*}

We first show in Figure \ref{fig:flux} that the measured flux of the continuum emission measured on the data presented by \cite{2019Manara+} agrees with the total flux measured on the new data within the 10\% uncertainty that is usually assumed on ALMA flux measurements. Exception to this good agreement are the circumsecondary discs around the stars in UZ Tau systems, which fluxes differ by $\sim 20-30\%$. 
    
We then compare the measured sizes of the discs on the old and new data, accounting for the differences in resolution and method.
Images of interferometric data are always the result of the convolution of the effective beam of the interferometer with the intensity profile of the science target, i.e., the disc. The size of the beam is thus a bigger effect when it is close to the physical size of the target.
In order to estimate the impact of the beam on our measurements, we first neglect the fact that the beam is elliptical with beam major axis BMAJ and minor axis BMIN, and we make the assumption that the beam can be modelled as a one-dimensional Gaussian:

\begin{equation}
    \mathcal{B}(x)=\frac{1}{\sqrt{2 \pi}\sigma_\mathrm{res}}\exp \biggl( - \frac{1}{2} \frac{(x-\mu)}{\sigma_\mathrm{res}^2} \biggr),
\end{equation}
where $\mu$ is the source centre and the standard deviation $\sigma_\mathrm{res}$ is linked to the beam size $\Gamma$ (FWHM beam) as
\begin{equation}
    \sigma_\mathrm{res} = \frac{\Gamma}{2 \sqrt{2 \ln{2}}}.
\end{equation}

The convolution of the beams size with the intrinsic (``true'') intensity profile of the disc is then calculated under the corresponding assumption that the disc can be described with a one-dimensional Gaussian. Hence, the standard  deviation $\sigma_\mathrm{true}$ of the Gaussian is equal to the effective disc radius $R_{68,\mathrm{true}}$ (since the effective disc radius is defined as the radius containing the $68\%$ of the total flux). The convolution between these two Gaussians is a one-dimensional Gaussian with standard deviation equal to
\begin{equation}
\sigma_\mathrm{obs}^2=\sigma_\mathrm{true}^2+\sigma_\mathrm{res}^2=R_{68,\mathrm{true}}^2+\sigma_\mathrm{res}^2,
\end{equation}
which represents the observed effective disc radius:  $\sigma_\mathrm{obs}\equiv R_{68,\mathrm{obs}}$.

\begin{table}[h]
\centering 
\small
\begin{tabular}{l*{2}{c}}

 &  Beam size [ $\arcsec\times\arcsec$] \\
\hline\hline
CIDA 9 &  $0.130 \times 0.099$ \\
DH Tau &  $0.132 \times 0.107$  \\
DK Tau &  $0.129 \times 0.107$ \\
HK Tau  &  $0.122 \times 0.107$ \\
HN Tau &  $0.142 \times 0.100$ \\
RW Aur &  $0.158 \times 0.100$ \\
T Tau &   $0.138 \times 0.100$   \\
UY Aur  &   $0.152 \times 0.099$  \\
UZ Tau &  $0.127 \times 0.105$ \\
V710 Tau &  $0.139 \times 0.099$ \\

\hline
\end{tabular}
\caption{Beam sizes of the observations by \cite{2019Manara+}.}\label{tab:beamOLD}
\end{table}

The data analyzed here have a spatial resolution of $0\farcs 15$, whereas the observations analysed by \cite{2019Manara+} had an angular resolution of $0\farcs12$. Tables \ref{tab:rms_beam_cont} and \ref{tab:beamOLD} show for each target the synthesised beams of the observations respectively for the new data and the data by \cite{2019Manara+}. 
 Since we are assuming a one-dimensional Gaussian intensity profile and beam, we are neglecting the position angle effect both of the disc and of the beam. Hence, the minimum BMIN gives an estimate of the best beam resolution, while the maximum BMAJ gives an estimate of the worst beam resolution. 
In the continuum data by \cite{2019Manara+}, the minimum BMIN and the maximum BMAJ are respectively $\Gamma_\mathrm{min,old}=0\farcs099$ and $\Gamma_\mathrm{max, old}=0\farcs142$ (BMIN of CIDA 9 and BMAJ of HN Tau, respectively). In the new continuum data, the minimum BMIN and the maximum BMAJ are respectively $\Gamma_\mathrm{min, new}=0\farcs120$ and $\Gamma_\mathrm{max, new}=0\farcs282$ (BMIN of HN Tau and BMAJ of V710 Tau, respectively).
In this way, the maximum effect on the ratio of the beam size between the data by \cite{2019Manara+} and the new data is given by
\begin{equation}\label{eq:beam1}
    \frac{\sigma_\mathrm{obs, new}}{\sigma_\mathrm{obs, old}} \biggl |_\mathrm{max} (\sigma_\mathrm{true})=\sqrt{\frac{\sigma_\mathrm{true}^2+\bigl[\Gamma_\mathrm{max, new}/(2\sqrt{2\ln{2}})\bigr]^2}{\sigma_\mathrm{true}^2+\bigr[\Gamma_\mathrm{min, old}/(2\sqrt{2\ln{2}})\bigr]^2}},
\end{equation}
 where $\sigma_\mathrm{true}$ is the ``true'' effective disc radius, i.e. $R_{68,\mathrm{true}}$.
 On the contrast, the minimum effect is given by
 \begin{equation}\label{eq:beam2}
    \frac{\sigma_\mathrm{obs, new}}{\sigma_\mathrm{obs, old}} \biggl |_\mathrm{min} (\sigma_\mathrm{true})=\sqrt{\frac{\sigma_\mathrm{true}^2+\bigl[\Gamma_\mathrm{min, new}/(2\sqrt{2\ln{2}})\bigr]^2}{\sigma_\mathrm{true}^2+\bigr[\Gamma_\mathrm{max, old}/(2\sqrt{2\ln{2}})\bigr]^2}}.
\end{equation}

If the disc is described by a one-dimensional Gaussian intensity profile, since $R_{95,\mathrm{true}}$ is defined as the radius containing the $95\%$ of the total flux, $R_{95} \equiv 2 R_{68} $. Therefore, when estimating the maximum/minimum effect of the difference in the beam size as a function of the disc radii $R_{95}$ the factor two simplifies, and Equation \eqref{eq:beam1} and \eqref{eq:beam2} hold with $\sigma_\mathrm{true}$ expressing the dust disc radii $R_{95}$.

\begin{figure*}[]
    	\centering
	    \includegraphics[width=\textwidth]{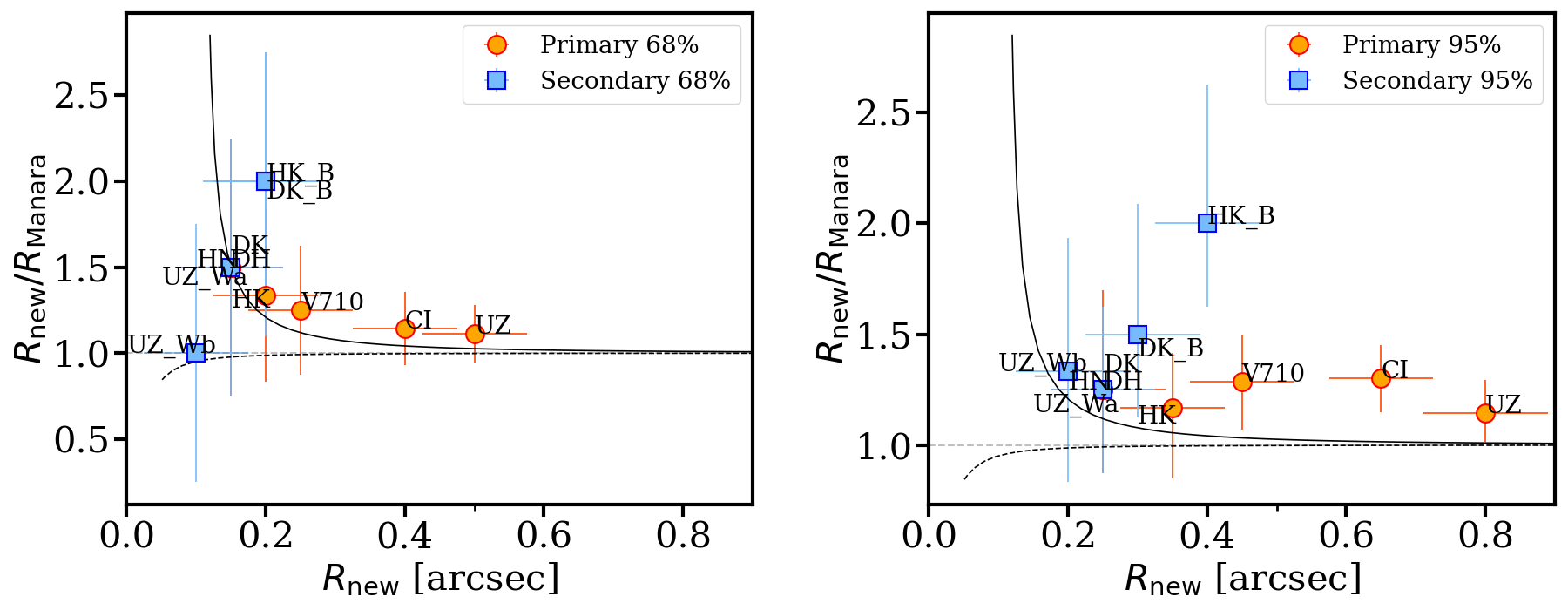}
	\caption{Observed ratio between the dust radii estimated with the cumulative flux technique applied on the data by \cite{2019Manara+} and the dust radii from the new continuum data ($R_{\mathrm{new}} / R_{\mathrm{Manara}}$) as a function of the new dust radii ($R_{\mathrm{new}}$). Left: dust radii estimated as the radii including the 68\% of the total flux (i.e. effective disc radii). Right: dust radii estimated as the radii including the 95\% of the total flux. Black lines show the maximum (solid) and minimum (dashed) effect on the ratio of the beam size between the data by \cite{2019Manara+} and the new data (see Equations \eqref{eq:beam1} and \eqref{eq:beam2}). Blue squares refer to the circumsecondary disc and red circles to the circumprimary discs.}
\label{fig:DUSTcfr}
\end{figure*}

Figure \ref{fig:DUSTcfr} shows these maximum (black solid line) and minimum (black dashed line) effects and compares these estimates with the observed ratio between the dust effective radii ($R_{68}$, left panel) and the dust radii ($R_{95}$, right panel) estimated with the cumulative flux technique applied on the data by \cite{2019Manara+} and the new continuum data. In general, as shown by the black lines, the smaller the disc size is, the greater the beam-size effect is, while, as expected, for larger discs the effect due to the different beam size tends towards 1.
The general trend of the observed ratio (red circles for circumprimary discs and blue squares for circumsecondary discs) shows a good agreement with the estimates (black lines). Figure \ref{fig:DUSTcfr} shows also a systematic shift on the $y$-direction of the observed ratio both between $R_{68}$ and $R_{95}$. We would expect such a systematic shift in the ratio between the dust radii $R_{95}$ (and not between $R_{68}$), due to the difference in integration time between the observations by \cite{2019Manara+} and the new observations $- \sim 4-9$ min/target versus $\sim 40$ min/target, respectively. This difference makes the new observations deeper, i.e. more sensitive, than the ones by \cite{2019Manara+}. However, as mentioned, Figure \ref{fig:flux} shows that the dust fluxes estimated on the new data and on the data by \cite{2019Manara+} are in good agreement with the 10\% uncertainty usually assumed on ALMA fluxes measurements. Therefore, in most systems, this systematic shift probably is not due to a sensitivity effect and the reason why it is observed must be further investigated. In particular, the systematic shift might be due to the assumption on the intensity profile of the disc, which actually is not a Gaussian, but a profile with less peaked emission in the centre, meaning that the 68\% (and 95\%) of the emission is distributed over a larger radius.

\begin{figure*}[]
    	\centering
	    \includegraphics[width=\textwidth]{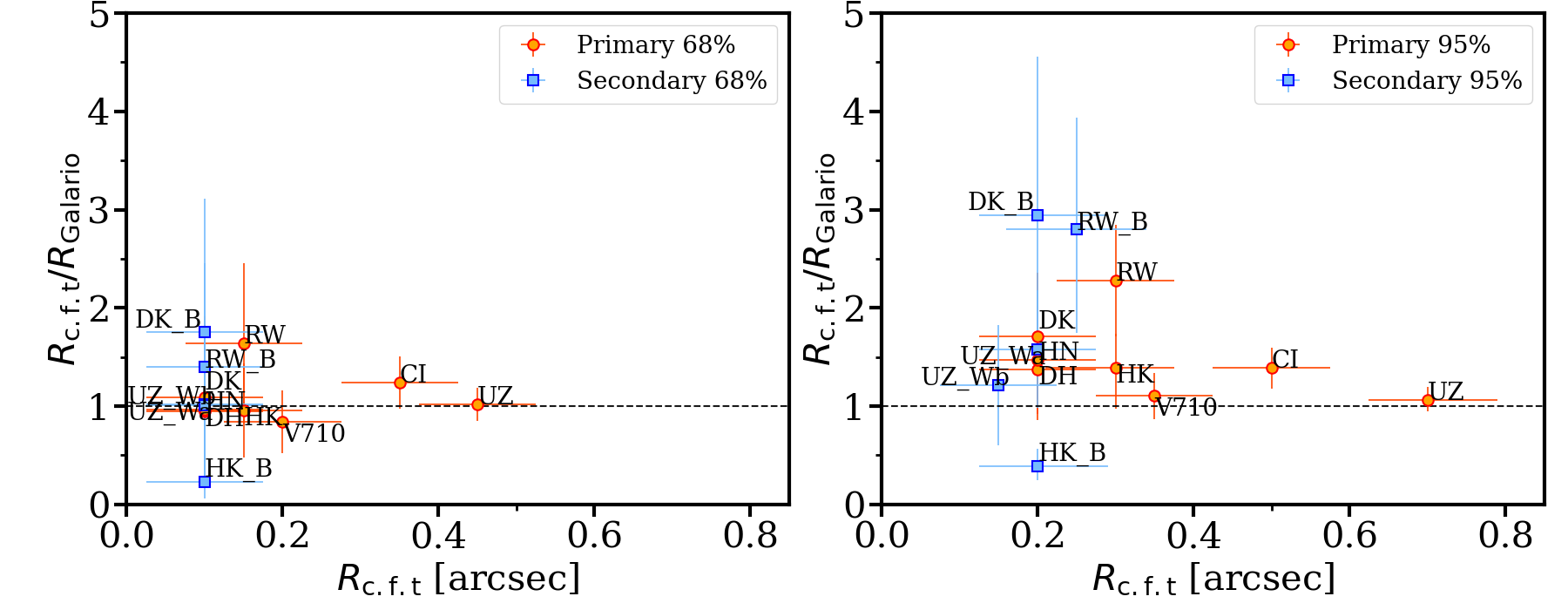}
	\caption{Comparison between dust disc radii estimated with the cumulative flux technique (c.f.t.) applied on the data by \cite{2019Manara+} and with those estimated by \cite{2019Manara+} with \textit{Galario}-modelling ($R_{\mathrm{c.f.t}} / R_{\mathrm{Galario}} $). Left: dust radii estimated as the radii including the 68\% of the total flux (i.e. effective disc radii). Right: dust radii estimated as the radii including the 95\% of the total flux.  Red circles are used for circumprimary discs, blue square for circumsecondary discs. }\label{fig:DUSTcfrold}
\end{figure*}

Finally, we compare the radii estimated with the analysis carried out in the $(u,v)$ plane \citep{2019Manara+} with those obtained with the method used in this work (Sect.~\ref{sect:analysis}). This is shown in Figure \ref{fig:DUSTcfrold}.
With the visibility plane analysis a resolution down to one third of the beam can be achieved \citep[e.g.,][]{2020Jennings+}. On the other hand, the cumulative flux technique is performed in the image plane,
thus the convolution of the beam with the intensity profile of the disc places an intrinsic resolution limit.

The left-hand panel in Figure \ref{fig:DUSTcfrold} shows the comparison between the effective dust radii $R_{68}$ estimated with the two methods. We cannot identify any particular difference between the two radii estimates, with the only exception being RW Aur discs and DK Tau B disc. We think that the fit with \textit{Galario} for RW Aur and the fit with \textit{eddy} for DK Tau B could have led to wrong results.

On the contrary, in the right panel showing the comparison between the dust radii $R_{95}$ it is observed that the dust radii estimated from the new observations are typically slightly larger than those obtained with the data by \cite{2019Manara+}. This is due to the cumulative flux technique, that gets the really outer extent of the discs, where little emission is detected (see Figure \ref{fig:mom0_radii}).

Since in the visibility plane analysis a resolution under the beam size is achievable, the dust radii estimated using \textit{Galario} are more reliable, especially for the more compact discs.

\end{document}